\documentclass[12pt]{article}
\usepackage{setspace}
\usepackage{float}
\usepackage{silence}
\usepackage{enumitem}
\usepackage{graphicx, epstopdf}
\usepackage{xparse,mathtools}
\usepackage{graphicx}
\usepackage{caption}
\usepackage{subcaption}
\usepackage{amsmath,amsfonts,amssymb}
\usepackage{amsthm}
\usepackage{breqn}
\usepackage[margin=1in]{geometry}
\usepackage{algorithm,algorithmic}
\usepackage{setspace}
\usepackage{color}
\usepackage{bm}
\usepackage{bbm}
\usepackage{multirow}
\usepackage{array}
\usepackage{natbib}
\usepackage{soul}
\usepackage[usenames,dvipsnames]{xcolor}
\usepackage{xr-hyper}

\graphicspath{ {./figures/}{./figures/bic/}{./}{./figures/ari_2d}{./figures/ari_additional}}

\usepackage{tikz}  
\usetikzlibrary{positioning}

\DeclareMathOperator*{\argmin}{arg\,min}
\newcommand{\norm}[1]{\left\lVert#1\right\rVert}

\allowdisplaybreaks

\ExplSyntaxOn

\NewDocumentCommand \vect { s o m }
{
	\IfBooleanTF {#1}
	{ \vectaux*{#3} }
	{ \IfValueTF {#2} { \vectaux[#2]{#3} } { \vectaux{#3} } }
}

\DeclarePairedDelimiterX \vectaux [1] {\lbrack} {\rbrack}
{ \, \dbacc_vect:n { #1 } \, }

\cs_new_protected:Npn \dbacc_vect:n #1
{
	\seq_set_split:Nnn \l_tmpa_seq { , } { #1 }
	\seq_use:Nn \l_tmpa_seq { \enspace }
}
\ExplSyntaxOff

\newcommand{\thresh}{0_95}
\newcommand{\arisize}{400}

\newcommand{\simonebicwid}{1.2in}

\newcommand{\gvec}[1]{\bm{#1}}

\makeatletter
\newcommand*{\addFileDependency}[1]{
  \typeout{(#1)}
  \@addtofilelist{#1}
  \IfFileExists{#1}{}{\typeout{No file #1.}}
}
\makeatother

\newcommand*{\myexternaldocument}[1]{%
    \externaldocument{#1}%
    \addFileDependency{#1.tex}%
    \addFileDependency{#1.aux}%
}

\myexternaldocument{FRkmeans_Supplementary}

\usepackage{hyperref}
\hypersetup{
    colorlinks = true,
    citecolor = {blue},
}

\title{Elastic $k$-means clustering of functional data for posterior exploration, with an application to inference on acute respiratory infection dynamics}
\author{Xiao Zang, Sebastian Kurtek, Oksana Chkrebtii, J. Derek Tucker}
\usepackage{authblk}
\author[1]{Xiao Zang}
\author[1]{Sebastian Kurtek}
\author[1]{Oksana Chkrebtii}
\author[2]{J. Derek Tucker}
\affil[1]{Department of Statistics, The Ohio State University}
\affil[2]{Statistical Sciences, Sandia National Laboratories}

\date{}

\begin{document}

\maketitle

\begin{abstract}
    We propose a new method for clustering of functional data using a $k$-means framework. 
    We work within the elastic functional data analysis framework, which allows for decomposition of the overall variation in functional data into amplitude and phase components. We use the amplitude component to  partition functions into shape clusters using an automated approach. To select an appropriate number of clusters, we additionally propose a novel Bayesian Information Criterion defined using a mixture model on principal components estimated using functional Principal Component Analysis. The proposed method is motivated by the problem of posterior exploration, wherein samples obtained from Markov chain Monte Carlo algorithms are naturally represented as functions. We evaluate our approach using a simulated dataset, and apply it to a study of acute respiratory infection dynamics in San Luis Potos\'{i}, Mexico.
\end{abstract}



\section{Introduction}

Bayesian inference is central in many modern applications including geological engineering \citep{IglesiasmEtAl2014}, biochemical kinetics \citep{Wilkinson2011}, studying the patterns of animal movement \citep{McDermottEtAl2017}, and many more. An important benefit of the Bayesian approach is that it flexibly accounts for uncertainty from different sources, where subjective belief about a set of unknown model components conditioned on observed data is expressed via posterior probabilities. Posterior uncertainty is often highly structured, such as when probable values of the unknown object lie on a low-dimensional manifold, exhibit complex correlation structure, or form multiple disjoint clusters. Processing and visualizing such structures is critical to assessing our understanding of unknown model components, be they vectors of model parameters, functions over time, or surfaces and shapes.

When inference is to be made on a limited number of scalar parameters and the probability mass of the posterior distribution is concentrated within a tractable domain, direct visualization could be achieved by density plots and contour plots using a grid approximation \citep{J_Kruschke_2014,A_Gelman_2013}. Most often, posterior samples are drawn by Markov chain Monte Carlo (MCMC) methods, and visualization is then done either through kernel density estimates (see, e.g., \citet{D_Rasmussen_2014} and \citet{RA_Rosales_2004}) or via standard graphical representations of univariate and multivariate data including one- and two-dimensional histograms and scatter plots (see, e.g., \citet{G_Zhu_2018}, \citet{I_Eisenkolb_2019} and \citet{A_Baetica_2016}). These visualization techniques help identify correlations and clusters in the samples. In addition, summaries such as the mean or quantiles, which can convey useful information about the posterior distribution such as its center and spread, are frequently estimated by the corresponding statistics computed from MCMC samples.

For higher-dimensional parameter spaces, visualization becomes more difficult. Common practice adopted in the Bayesian analysis literature is to apply the aforementioned visualization tools marginally, usually up to two dimensions at a time. However, it has been pointed out that this approach may fail to identify structures such as multimodality \citep{J_Venna_2003}. One can foresee that exploration of vector-valued MCMC samples can potentially benefit from more sophisticated techniques for visualizing multivariate data, many of which are surveyed in \citet{S_Liu_2017} and \citet{G_Grinstein_2001}. But, in modern application scenarios, the inferential object is often a function, such as the temporal evolution of a variable or an unknown initial condition for a partial differential equation model, where the infinite-dimensionality of the object under study poses an even greater challenge for posterior visualization and summarization. Unlike the large repertoire of methods for multivariate data, the techniques to visualize a sample of functions are relatively limited. One possible approach is to map the data into a finite-dimensional space using dimensionality reduction such as functional Principal Component Analysis (fPCA), and analyze the resulting fPCA coefficient vectors. Visualizations can then be mapped back to the original space to provide better interpretation in terms of the original functions \citep{R_Hyndman_2010, tucker2013}. Instead of resorting to dimension reduction, \citet{Y_Sun_2011} construct functional boxplots using the notion of data depth for functional observations \citep{S_Lopez_2009}. An alternative boxplot-type graphic that decomposes variability in observed functions into various components (not related to fPCA) was proposed by \citet{X_Weiyi_2017}.

\begin{figure}[!t]
	\begin{center}
		\begin{tabular}{|c c|c c|}
		\hline
		\multicolumn{2}{|c|}{(a)}&\multicolumn{2}{|c|}{(b)}\\
		\hline
		\includegraphics[width=1.4in]{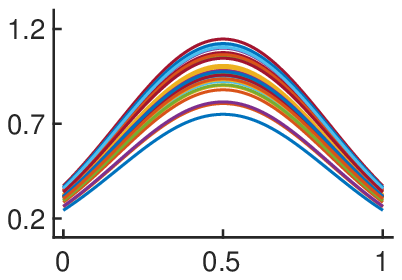} &
			\includegraphics[width=1.4in]{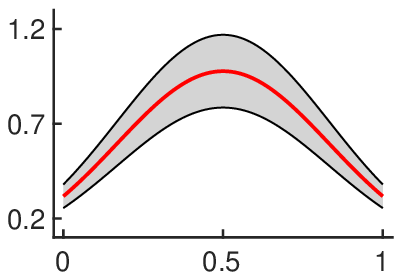} &
			\includegraphics[width=1.4in]{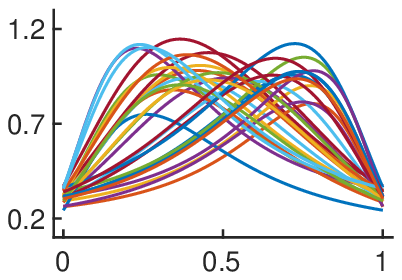} &
			\includegraphics[width=1.4in]{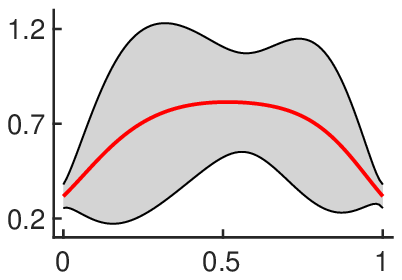}\\
		\hline
		\end{tabular}
		\caption{Pointwise summaries of misaligned functional data have distorted shapes and result in overestimated amplitude variability. Sample of (a) aligned and (b) misaligned functions, and the resulting pointwise two standard deviation bands around the mean.}
		\label{fig:example_ptsumproblem}
	\end{center}
\end{figure}

When dealing with samples of functions, even simple measures of center and dispersion are not trivial to obtain. For example, pointwise means, quantiles and standard deviations, in spite of being prevalent approaches to summarize posterior sample functions, suffer from a commonly encountered issue in functional data analysis: the misalignment of features such as peaks and valleys. This is caused by two sources of variability often jointly contributing to the overall variation of observed functions: vertical variability along the $y$-axis referred to as amplitude variability, and the lateral displacement along the $x$-axis called phase variability \citep{J_Marron_2015}. The latter could be alternatively viewed as the variability induced by domain transformations or ``warpings'', and the amplitude component could then be intuitively interpreted as the shape (and scale) of a function that is invariant under domain warpings. In rare cases where there is negligible phase variability, as depicted in Figure \ref{fig:example_ptsumproblem}(a), the pointwise two standard deviation band around the mean provides a good characterization of the typical shape and scale of the functions in the sample. On the other hand, in the presence of substantial phase variability, as shown in Figure \ref{fig:example_ptsumproblem}(b), the same pointwise summaries have highly distorted shapes; furthermore, scale variability is exaggerated, resulting in a significantly wider band. Additional illustrative examples of pitfalls caused by misalignment in functional data samples can be found in \citet{A_Srivastava_2011}. 

\subsection{Motivating Application: Inference on Infection States of Acute Respiratory Infections}

Acute respiratory infections (ARI) are a public health concern around the world. They include illness caused by viruses such as influenza, respiratory syncytial virus (RSV), parainfluenza, rhinovirus, and the recently discovered coronavirus that leads to COVID-19. The circulation of different strains of ARI-causing pathogens each year makes future years' patterns difficult to predict. However, fitting circulation models to past data can yield important information about (i) individual characteristics of a specific pathogen strain, such as transmission rates, and (ii) characteristics of the interaction between pathogens circulating in the same year, such as cross-immunity, and importantly, (iii) the dynamics of the epidemic, such as the occurrence of epidemic peaks and their relative order during the year. The first two features correspond to multivariate inputs in commonly-used circulation models, and thus posterior visualization can be conducted using existing approaches. On the other hand, the time-evolution of infections has a functional structure and requires the development of appropriate posterior visualization techniques.

\begin{figure}[!t]
	\begin{center}
		\begin{tabular}{|c|c|c c|}
			\hline
			& (a) & (b) & (c) \\
			\hline
			(1) &
			\includegraphics{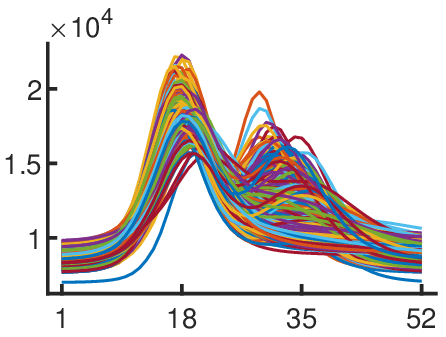} &
			\includegraphics{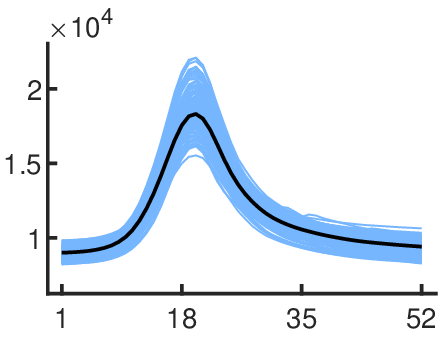} &
			\includegraphics{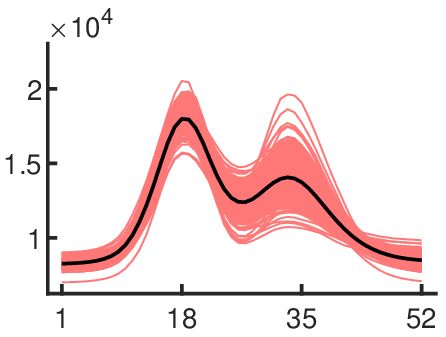} \\
			\hline
			(2) &
			\includegraphics{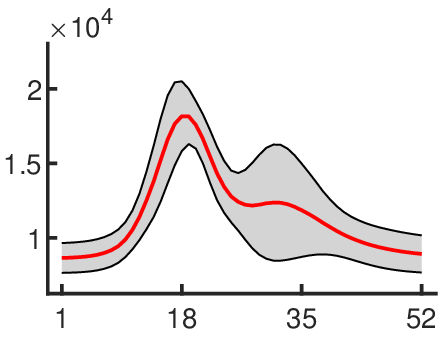} &
			\includegraphics{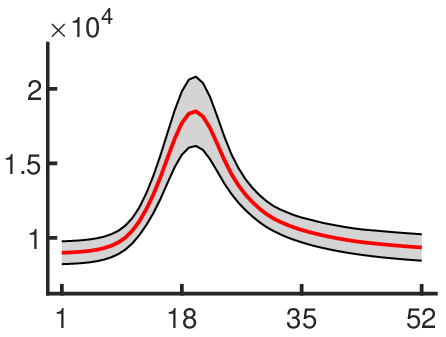} &
			\includegraphics{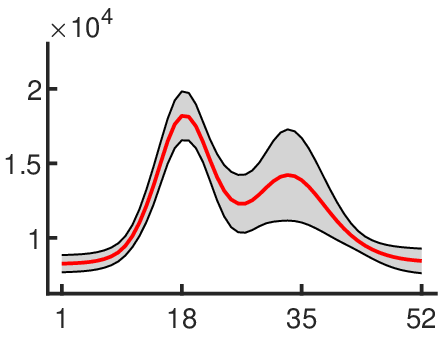} \\
			\hline
		\end{tabular}
		\caption{Application of the proposed elastic $k$-means clustering to 400 posterior draws of aggregated ARI trajectories for the year 2002-03.  (1)(a) Original sample of functions. (1)(b)-(c) Two clusters (single mode cluster in blue and cluster with two modes in red) of functions that were optimally aligned within each cluster (solid black functions are estimated cluster templates). The number of clusters was automatically determined via a proposed modification of the Bayesian Information Criterion (BIC). (2)(a)-(c) Pointwise summaries (mean $\pm$ two standard deviations) for the functions shown in (1)(a)-(c), respectively.}
		\label{fig:ari_2002}
	\end{center}
\end{figure}

A popular model structure for the evolution of infection states is the Susceptible-Infected-Recovered (SIR) framework which describes transitions between infection states. Although a deterministic compartmental model can approximate reality well in some situations, stochastic SIR modeling better reflects the stochastic nature of disease transmission events \citep{L_Allen_2017}. Such stochastic kinetic models (SKM) define a distribution over infection states associated with a given parameter setting, so that the marginal posterior distribution over infection states cannot be directly recovered from the marginal posterior over the model parameters, as would be the case under a deterministic SIR model. Furthermore, a common feature of ARI is that, although they are readily identified by a physician, the ARI-causing pathogen cannot be determined based on symptoms alone. Due to cost considerations, tests for the ARI-causing pathogens are not routinely conducted and thus only aggregated data is available, except for small at-risk subsets of the population. This problem structure introduces the potential of different underlying epidemic dynamics leading to similar observed patterns of infection, which manifests in posterior multimodality over the evolution of infection states. This behavior is illustrated in the top left panel of Figure \ref{fig:ari_2002}, which displays 400 functions sampled from the posterior distribution of the aggregated mean infection trajectories of influenza and RSV, in the epidemic year 2002-03, for data collected in San Luis Potos\'{i}, Mexico. More information about the model, data and statistical analysis is provided in \citet{Y_Garcia_2017}. It seems possible that the posterior sample is made up of at least two qualitatively different outbreak patterns, one which has a single peak, and one that has two. These patterns, however, are not clearly discernible in the spaghetti plot due to the presence of overlapping misaligned functions. The lower panel of Figure \ref{fig:ari_2002}(a) shows pointwise summaries of the infection trajectories, directly applied to the posterior sample (top of panel (a)), where the possible presence of multiple posterior clusters is obscured, and a moderate level of distortion is present due to misalignment. The result given by our proposed approach, which is outlined in subsequent sections, is shown in panels (b) and (c). We identify two distinct clusters in the posterior sample corresponding to different temporal infection patterns. An additional benefit of such clustering is that alignment within each cluster results in improved pointwise summarization of the sample.  

More generally, similar issues arise for Bayesian models with function-valued unknowns that are not fully identified given the data. Therefore, it is important to devise a systematic analysis technique for functional posterior samples that is able to identify and distinguish disjoint clusters of trajectories while also resolving misalignment.

\subsection{Contributions}
\label{sec:contr}

For a sample of functions, such as ones generated via MCMC, we aim to achieve the following objectives prior to summarization and visualization of variability:

\begin{enumerate}[nosep, topsep=-1em]
    \item Identify distinct functional shapes within the sample;
    \item Partition the sample into such shape clusters using an automated approach; and
    \item Align functions within each cluster to separate amplitude and phase variabilities.
\end{enumerate}
Visualization and pointwise summarization within each shape cluster provides a natural strategy for exploring posterior structure and variability. With this in mind, the present paper makes the following contributions to the literature:
\begin{enumerate}[nosep, topsep=-1em]
\item We introduce a novel algorithm, referred to as elastic $k$-means, for clustering functions based on their shape;
\item We propose a criterion-based approach for automatic selection of the number of clusters in a sample of functions;
\item We employ the proposed methodology to interpret the results of a Bayesian analysis of the time evolution of acute respiratory infections in San Luis Potos\'{i}, Mexico from \citet{Y_Garcia_2017}.
\end{enumerate}
The performance of the proposed approach is also evaluated via multiple simulation studies. 

The rest of the paper is organized as follows. Section \ref{sec:background} briefly reviews background material on elastic functional data analysis and discusses existing methods for functional data clustering. Section \ref{sec:elastickmeans} outlines the proposed elastic $k$-means approach and defines a model-based criterion for automatic selection of the number of clusters. Section \ref{sec:simulations} describes simulation studies that validate the proposed approach and compare its performance to that of other applicable methods. The application of our approach to the aforementioned ARI posterior samples is presented in Section \ref{sec:application}. Section \ref{sec:discussion} provides a short discussion and describes directions for future work. 
\section{Background Material}
\label{sec:background}

We begin with a brief description of function alignment, and a technical review of existing methods for clustering of functional data. 

\subsection{Elastic Function Alignment Without Clustering}
Under the assumption that no clustering structure exists in a sample of functions, the three objectives outlined in Section \ref{sec:contr} reduce to the single task of function alignment, also commonly referred to as registration. Thus, we begin by formally defining the problem of function alignment, and highlight a framework that is later used to define the proposed elastic $k$-means clustering approach. 

\subsubsection{Pairwise Alignment} \label{sec:pairalign}

Given two functions $f_1$ and $f_2$ belonging to a function space $\mathcal{F}$ equipped with a distance metric $d(\cdot, \cdot)$, pairwise alignment seeks an optimal warping $\gamma^*$, within some set of warping functions $\Gamma$, that minimizes this distance, i.e., $\gamma^* =\argmin_{\gamma \in \Gamma} d(f_1, f_2 \circ \gamma)$, where $f \circ \gamma$ denotes function composition. For the alignment problem to be well-defined, the triple $(\mathcal{F}, d, \Gamma)$ must satisfy the isometry property: for any $f_1, f_2 \in \mathcal{F}$ and $\gamma \in \Gamma$, $d(f_1, f_2) = d(f_1 \circ \gamma, f_2 \circ \gamma)$. That is, the distance between two functions should be unchanged by simultaneous warping. A common approach in the literature uses $\mathcal{F} = \mathbb{L}^2([0,1])$, $d(f_1, f_2) = \norm{f_1 - f_2} \coloneqq \sqrt{\int_0^1 |f_1 - f_2|^2 dt}$, and $\Gamma = \{ \gamma : [0,1] \rightarrow [0,1] \mid \gamma(0) = 0, \gamma(1) = 1, \dot{\gamma} > 0\}$, where $\dot{\gamma}$ is the derivative of $\gamma$.  Unfortunately, these choices do not result in isometry under warping, resulting in the so-called pinching effect and registration asymmetry \citep{J_Marron_2015,A_Srivastava_2016}. Examples of methods that use the $\mathbb{L}^2$ metric for registration include \citet{JO_Ramsay_1998} and \citet{X_Liu_2004}.

Alternatively, \citet*{A_Srivastava_2011_b} and \citet*{A_Srivastava_2011} consider $\mathcal{F} = \{f:[0,1] \rightarrow \mathbb{R}^m \mid f \text{ is absolutely continuous}\}$ equipped with an elastic Riemannian metric closely related to the well-known Fisher-Rao (FR) metric for probability density functions \citep{C_Rao_1945}, and $\Gamma$ defined the same as in the previous paragraph. While these choices result in isometry under warping, the resulting distance is difficult to use in practice and requires computationally intensive numerical algorithms for registration \citep{W_Mio_2007}. Instead, one can simplify the problem via a simple transformation of the original data as follows. Define the square-root velocity function (SRVF) as $q \coloneqq \frac{\dot{f}}{\sqrt{|\dot{f}|}}$ where $|\:\cdot\:|$ denotes the $\ell^2$-norm ($q \coloneqq 0$ when $|\dot{f}| = 0$). The space of SRVFs, corresponding to $\mathcal{F}$, is $\mathcal{Q}=\mathbb{L}^2([0, 1])$, and the aforementioned complicated elastic Riemannian metric on $\mathcal{F}$ simplifies to the simple $\mathbb{L}^2$ metric on $\mathcal{Q}$, also retaining all of its theoretical properties. Consequently, the distance between two functions $f_1$ and $f_2$ can be computed via the $\mathbb{L}^2$ distance between their SRVFs $q_1$ and $q_2$. A warping of a function $f$, $f\circ\gamma$, results in a transformation of its SRVF $q$ given by $(q,\gamma)=(q \circ \gamma) \sqrt{\dot{\gamma}}$. 
Let $[q]=\{(q,\gamma)|\gamma\in\Gamma\}$ denote the orbit of an SRVF under warping. Pairwise registration refers to minimizing $\norm{q_1 - (q_2, \gamma)}$ over $\gamma \in \Gamma$, which leads to a distance between two orbits, $[q_1],\ [q_2]$, defined as $d([q_1], [q_2]) \coloneqq \inf_{\gamma \in \Gamma} \norm{q_1 - (q_2, \gamma)}$. Such a distance provides a measure of similarity between the shapes of two functions, i.e., the orbit defines the function's amplitude and the distance is then the amplitude distance. The optimal warping $\gamma^*$ can be identified using the dynamic programming algorithm \citep{D_Robinson_2012} or Riemannian optimization \citep{huang2016}. We use $f_2^* = f_2 \circ \gamma^*$ to denote the function $f_2$ after optimally aligning it to $f_1$ (corresponding to $q_2^* = (q_2, \gamma^*)$).

\subsubsection{Multiple Alignment} \label{sec:multalign}

Multiple alignment is generally formulated \emph{with respect to a template}, usually defined as the sample mean. Let $f_1,\dots,f_N$ denote the observed functions, and $q_1,\dots,q_N$ be their SRVFs. Then, under the SRVF representation, the sample mean amplitude orbit is defined via the amplitude distance as $[\bar{q}] \coloneqq \argmin_{[q] \in \mathcal{Q}/\Gamma}\sum_{i=1}^N d^2([q], [q_i])$; note that the amplitude distance involves pairwise registration of each function in the sample to the estimated template. In practice, one must choose a single representative element in this orbit, $\bar{q}\in[\bar{q}]$. This is done via a centering step, which ensures that the average of warping functions estimated using pairwise alignment of the data to $\bar{q}$ is the identity warping $\gamma_{id}(t)=t$ \citep{A_Srivastava_2011}. Thus, the algorithm for multiple alignment can be viewed as a two step process: (1) estimate the sample mean amplitude and select a representative element in its orbit, and (2) perform pairwise alignment of each function in the data to this representative element. The entire procedure results in the template $\bar{f}$ (corresponding to $\bar{q}$), optimal warping functions \{$\gamma_1^*,\dots,\gamma_N^*$\}, and registered functions \{$f_1^*=f_1\circ\gamma_1^*,\dots,f_N^*=f_N\circ\gamma_N^*$\} (corresponding to \{$q_1^*=(q_1,\gamma_1^*),\dots,q_N^*=(q_N,\gamma_N^*)$\}). 

\subsection{Existing Methods for Clustering of Functional Data}

Clustering of functional data has received great interest in recent years, and many different methods have been developed for this task; see the survey by \citet{J_Jacques_2014} and references therein. The most naive approach is to view $f_i([t])$, the values of a function $f_i$ evaluated at $T$ points along its domain $[t]=(t_1,\dots,t_T)^\intercal$, as multivariate data. Then, one can directly apply standard methods for clustering vector-valued observations; many such methods exist in the literature \citep{D_Xu_2015, A_Saxena_2017}, including $k$-means clustering \citep{E_Forgy_1965}, hierarchical clustering \citep{J_Ward_1963}, Gaussian mixture models (GMM) \citep{J_Banfield_1993}, among others. However, treating discretized functions as vectors ignores the temporal dependence inherent in functional data. Such an approach is also inappropriate when the recorded time points are different across functions, which is the case in many applications. An alternative approach is to reduce the dimension by representing functional data using basis function coefficients. One example is \citet{C_Abraham_2003} who performed $k$-means clustering on B-spline basis coefficients estimated separately for each function via least squares. \citet{G_James_2003} also utilized basis functions, but they proposed a model-based framework assuming that the basis coefficients come from a GMM, with each $f_i([t])$ observed on a sparse grid of time points, conditioned on the basis coefficients, following a multivariate Gaussian distribution. Distance-based clustering methods generally do not rely on basis expansions, and can be easily applied to cluster functional data. An example is \citet{T_Tarpey_2003} who used the $\mathbb{L}^2$ distance in a functional $k$-means clustering algorithm. 

None of the aforementioned approaches consider separation of amplitude and phase variabilities in functional data, i.e., they assume that the data is already aligned or that phase variability is negligible. However, in many applications, as extensively evidenced in \cite{A_Srivastava_2016}, functional data analysis benefits from separation of amplitude and phase. 
Compared to the abundance of methods for clustering functional data without alignment, the literature for methods that find amplitude clusters through function alignment is relatively scarce. \citet{X_Liu_2009} include time shifts as incomplete data into model-based clustering, but applications of this method are limited due to a very simple warping structure. \citet{L_Sangalli_2010} present the first attempt to extend $k$-means clustering to the problem of finding amplitude clusters. The function space and distance metric considered in their paper are $\mathcal{F} = \{f \in \mathbb{L}^2(\mathbb{R})\mid \dot{f} \in \mathbb{L}^2(\mathbb{R}),\ |\dot{f}| \neq 0\}$ and $d(f_1, f_2) = \norm{\frac{\dot{f_1}}{\norm{\dot{f_1}}} - \frac{\dot{f_2}}{\norm{\dot{f_2}}}}$, respectively. In order to satisfy the isometry property under the chosen metric, their approach restricts warpings to the set of strictly increasing affine transformations: $\Gamma = \{\gamma: \mathbb{R} \rightarrow \mathbb{R} \mid \gamma = at + b,\ a \in \mathbb{R}^+,\ b \in \mathbb{R} \}$. Although this warping model allows shifts and dilations of the time domain, linear warpings are often not flexible enough in practice. Another issue is the need to define the domain of all functions to be the entire real line to develop the necessary theory. Practically, the functions are always only observed on a finite interval, resulting in computational issues during implementation. Given this choice of the triplet $(\mathcal{F}, d, \Gamma)$, the optimization problem for $k$-means amplitude clustering of $N$ functions into $K$ clusters is given by minimization of $L(\gvec{\mu}, \gvec{\delta}) \coloneqq \sum_{i = 1}^{N} d^2([\mu_{\delta_i}], [f_i])$ ($[f]$ is the orbit of $f$ under the action of $\Gamma$ on $\mathcal{F}$) with respect to cluster template functions $\gvec{\mu} \coloneqq (\mu_1(t),\dots, \mu_K(t))$ and cluster assignments $\gvec{\delta} \coloneqq (\delta_1,\dots, \delta_N)$, where $\delta_i \in \{1,\dots,K\}$. Since this approach is most closely related to the proposed elastic $k$-means algorithm, we use it as a state-of-the-art benchmark comparison in Section \ref{sec:simulations}.

\section{Elastic $k$-means Algorithm with Automatic Selection of Number of Clusters}
\label{sec:elastickmeans}

Clustering via $k$-means with elastic alignment, using the triplet $(\mathcal{F}, d, \Gamma)$ as specified by \citet{A_Srivastava_2016}, can be viewed as a generalization of multiple function alignment (Section \ref{sec:multalign}), by allowing $K$ different templates. Formally, we aim to minimize the cost function $L(\bm{\eta}, \bm{\delta})=\sum_{i=1}^N d^2([\eta_{\delta_i}], [q_i])$ over cluster templates $\eta_k$ (represented using the SRVF), $k = 1, ..., K$, and cluster assignments $\delta_i \in \{1,...,K\},\ i = 1,...,N$. Here, the distance is the $\mathbb{L}^2$ distance between SRVF orbits, as defined in Section \ref{sec:pairalign}. The proposed clustering algorithm is built in similar fashion to that of \citet{L_Sangalli_2010}. In particular, it iterates through four main steps: (1) alignment, (2) assignment of each function to a cluster, (3) centering within orbits for each cluster, and (4) estimation of cluster templates. The alignment step performs pairwise registration of each function in the given data to each cluster template using the elastic distance. The assignment step assigns each function to a cluster based on the minimum amplitude distance between that function and the cluster templates. As in multiple alignment, the centering step ensures that the average of warping functions estimated using pairwise alignment of the data within each cluster to the estimated cluster template is the identity warping $\gamma_{id}(t)=t$ \citep{A_Srivastava_2011}. Finally, template estimation for each cluster is straightforward, as it is based on the cross-sectional mean of the SRVFs assigned to each cluster. 

The details of the full elastic $k$-means algorithm are as follows. Here, we assume that the observed data $f_1,\dots,f_N$ was first transformed into the corresponding SRVFs $q_1,\dots,q_N$.
\begin{enumerate}[topsep=0pt]
	\item Set the initial cluster templates $\{\eta_k^{(0)},\ k=1,\dots,K\}$ to a random sample of size $K$ from the given SRVFs $\{q_i,\ i=1,\dots,N\}$, without replacement.
	\item For $n = 1,\dots,n_{max}$.
	\begin{enumerate}
		\item Alignment: For $k = 1,\dots, K$ and $i = 1,\dots,N$ align $q_i$ to $\eta_k^{(n-1)}$ to get optimal warping functions $\gamma_{ik}^{(n)*}$, aligned SRVFs $q_{ik}^{(n)*}$ and amplitude distances $d([q_i], [\eta_k^{(n-1)}])$ for all $i=1,\dots,N$ and $k=1,\dots,K$.
		\item Assignment to clusters: For $i = 1,\dots,N$, set the cluster index of the $i^{\text{th}}$ function as $\delta_i^{(n)} = \argmin_{k \in \{1,\dots,K\}} d([q_i], [\eta_k^{(n-1)}])$. Let $M_k^{(n)} \coloneqq \{i \in 1,\dots,N \mid \delta_i^{(n)}=k\}$ denote the set of the indices of functions assigned to the $k^{\text{th}}$ cluster.
		\item Orbit centering within each cluster using the method of \cite{A_Srivastava_2011}.
		\item Template estimation in each cluster: Update $\eta_k^{(n)}$ to be the cross-sectional mean of $\{q_{ik}^{(n)*}\}_{i \in M_k^{(n)}}$ for $k = 1,\dots,K$.
		\item Stop if $\frac{1}{K}\sum_{k=1}^{K}\frac{\|\eta_k^{(n)} - \eta_k^{(n-1)}\|}{\|\eta_k^{(n-1)}\|} < \epsilon$, $\epsilon>0$ and small. Otherwise, continue to the next iteration.
	\end{enumerate}
\end{enumerate}

In each iteration, Step 2(a) finds all pairwise amplitude distances between orbits of cluster templates and orbits of observed functions. Then, Step 2(b) minimizes $L(\bm{\eta}, \bm{\delta})$ with respect to $\bm{\delta}$. Step 2(c) does not change the value of the cost function, and simply ensures that the cluster templates are identifiable. At the end, Step 2(d) further reduces the cost function, because 
\begin{equation*}
\begin{split}
& L(\bm{\eta}^{(n-1)}, \bm{\delta}^{(n)}) = \sum_{i=1}^N d^2([\eta_{\delta_i^{(n)}}^{(n-1)}], [q_i])\\ &= \sum_{k=1}^K\sum_{i \in M_k^{(n)}} \|\eta_k^{(n-1)} - q_{ik}^{(n)*}\|^2 \geq \sum_{k=1}^K\sum_{i \in M_k^{(n)}} \|\frac{1}{|M_k^{(n)}|}\sum_{j \in M_k^{(n)}}q_{jk}^{(n)*} - q_{ik}^{(n)*}\|^2 \\ &= \sum_{k=1}^K\sum_{i \in M_k^{(n)}} \| \eta_k^{(n)} - q_{ik}^{(n)*}\|^2\geq  \sum_{k=1}^K\sum_{i \in M_k^{(n)}} d^2([\eta_k^{(n)}],[q_i]) = L(\bm{\eta}^{(n)}, \bm{\delta}^{(n)}),
\end{split}
\end{equation*}
where $|M_k^{(n)}|$ denotes the number of functions in cluster $k$ after iteration $n$. Therefore, each full iteration, composed of Steps 2(a)-(d) decreases the cost function, which is bounded below by zero. Thus, the proposed elastic $k$-means algorithm is guaranteed to converge. To avoid empty clusters after Step 2(b), the proposed algorithm uses linear programming when determining cluster assignments \citep{P_Bradley_2000}.

\subsection{Model-based Estimation of the Number of Clusters}

The elastic $k$-means clustering method, as most other $k$-means algorithms, requires that the number of amplitude clusters $K$ is known \textit{a priori}. However, the choice of $K$ is difficult in most scenarios, including the motivating application to exploration of variability in Bayesian posterior sample functions. Thus, to alleviate this issue, we propose a model-based method for selecting $K$ based on the Bayesian Information Criterion (BIC). In particular, we use a combination of dimension reduction, a Gaussian mixture model (GMM) and the BIC. For clarity, we describe our approach for one-dimensional functional data only, and note that the case of higher-dimensional data can be handled with minor adjustments. 

We begin by applying elastic $k$-means clustering to the given data for $K \in \{1,...,K_{max}\}$, where $K_{max}$ is the maximum number of clusters; the choice of $K_{max}$ is user and application specific. For the result of elastic $k$-means clustering with the number of clusters set to $K$, let $q_{Ki}^*$ denote the SRVF of the $i^{\text{th}}$ function aligned to its corresponding cluster template, and $\delta_{Ki} \in \{1,...,K\}$ denote its cluster membership. Then, $M_{Kk} \coloneqq \{i \in 1,\dots,N \mid \delta_{Ki}=k\}$ is the set of indices of the functions belonging to the $k^{\text{th}}$ cluster. Within each cluster, we perform fPCA on the aligned functions (amplitude component) $\{q_{Ki}^*\}_{i \in M_{Kk}}$ as follows \citep{tucker2013}. Let $q_{Ki}^*([t]) = \left(q_{Ki}^*(t_1),\dots,q_{Ki}^*(t_T)\right)$ denote a vector (of size $1\times T$) of evaluations of the function $q_{Ki}^*$ at $T$ equally-spaced time points; the width of the time intervals used for discretization will be denoted by $\Delta t$. 
Then, $\textbf{Q}_{Kk}$ is a $\left|M_{Kk}\right| \times T$ matrix whose rows are given by $\left\{q_{Ki}^*([t]) - \frac{1}{|M_{Kk}|} \sum_{i \in M_{Kk}}q_{Ki}^*([t]) \middle| i \in M_{Kk} \right\}$. We apply singular value decomposition (SVD) to this data matrix,
$\textbf{Q}_{Kk} = \textbf{U}_{Kk}\bm{\Omega}_{Kk}\textbf{V}_{Kk}^T$, resulting in the orthogonal matrix 
$\textbf{U}_{Kk}$ of size $\left|M_{Kk}\right| \times \left|M_{Kk}\right|$, the diagonal matrix of non-negative entries $\bm{\Omega}_{Kk}$ of size $\left|M_{Kk}\right| \times T$, and the orthogonal matrix $\textbf{V}_{Kk}$ of size $T\times T$. Then, for $j \leq T$, the $j^{\text{th}}$ diagonal element of $\bm{\Lambda}_{Kk} = \frac{\Delta t}{\left|M_{Kk}\right|}\bm{\Omega}_{Kk}^T\bm{\Omega}_{Kk}$, denoted by $\lambda_{Kkj}$, is the variance of the $j^{\text{th}}$ PC. Also, the $j^{\text{th}}$ column of $\textbf{W}_{Kk} = \frac{1}{\sqrt{\Delta t}}\textbf{V}_{Kk}$, denoted by $\textbf{w}_{Kkj}$, is the discretized weight function corresponding to the $j^{\text{th}}$ PC. Then, the dimension $d$ (number of PCs) that is used for specifying the GMM and computing the BIC, is set to the minimum number of PCs needed to explain at least $(\rho\times 100)\%$ of variability in any of the clusters across all values of $K$. Note that this choice ensures that the dimension is the same across all $K=1,\dots,K_{max}$.

To reduce dimension, we represent each function using the vector of the first $d$ PC coefficients, $\textbf{c}_{Ki} \coloneqq \left(c_{Ki1},...,c_{Kid}\right)^T$, given by $c_{Kij} = \Delta t\ q_{Ki}^*([t]) \textbf{w}_{K\delta_{Ki}j} $ for $j = 1,\dots,d$. Then, for each choice of $K$, we view $\{(\delta_{Ki}, \textbf{c}_{Ki})\}$ as $iid$ observations from a Gaussian mixture model with $K$ components. The joint density is given by $f(\delta, \textbf{c}) = \alpha_{K\delta}(2\pi)^{-\frac{d}{2}}|\bm{\Sigma}_{K\delta}|^{-\frac{1}{2}}\exp\{-\frac{1}{2}(\textbf{c} - \bm{\mu}_{K\delta})^\intercal\bm{\Sigma}_{K\delta}^{-1}(\textbf{c} - \bm{\mu}_{K\delta})\}$, where $\alpha_{Kk}$, $\bm{\mu}_{Kk}$ and $\bm{\Sigma}_{Kk}$ are the mixture probability, and mean vector and covariance matrix in the $k^{\text{th}}$ cluster, respectively. The log-likelihood is given by 
\begin{align*}
&l(\alpha_{K1},...,\alpha_{KK}, \bm{\mu}_{K1}, ..., \bm{\mu}_{KK}, \bm{\Sigma}_{K1}, ...,\bm{\Sigma}_{KK}) =\\ &\sum_{i=1}^{N}\log(\alpha_{K\delta_i}) - \frac{1}{2}\sum_{i=1}^{N}\left( d\log\left(2\pi\right) + \log\left|\bm{\Sigma}_{K\delta_i}\right| + \left(\bm{c}_{Ki} - \bm{\mu}_{K\delta_i}\right)^\intercal\bm{\Sigma}_{K\delta_i}^{-1}\left(\textbf{c}_{Ki} - \bm{\mu}_{K\delta_i}\right) \right).
\end{align*}
Since PC coefficients are uncorrelated within each cluster, the clusterwise covariance matrices are diagonal. The Maximum Likelihood Estimates (MLEs) for all parameters are given by $\widehat{\alpha}_{Kk} = \frac{\left|M_{Kk}\right|}{N},\ \widehat{\bm{\mu}}_{Kk} = \frac{1}{\left|M_{Kk}\right|}\sum_{i \in M_{Kk}}\textbf{c}_{Ki},\ \widehat{\bm{\Sigma}}_{Kk} = \text{diag}\left(\widehat{\sigma}_{Kk1}^2,\dots, \widehat{\sigma}_{Kkd}^2 \right)$, where
$\widehat{\sigma}_{Kkj}^2 = \frac{1}{\left|M_{Kk}\right|}\sum_{i \in M_{Kk}}\left(c_{Kij} - \widehat{\mu}_{Kkj}\right)^2$ and  $\widehat{\mu}_{Kkj}$ is the $j^{\text{th}}$ element of $\widehat{\bm{\mu}}_{Kk}$. This results in $(2d + 1)K - 1$ estimated parameters. 
Therefore, the BIC for the number of clusters $K$, based on this GMM model, is given by  
\begin{equation}
BIC_K = -2l(\widehat{\alpha}_{K1},...,\widehat{\alpha}_{KK}, \widehat{\bm{\mu}}_{K1}, ..., \widehat{\bm{\mu}}_{KK}, \widehat{\bm{\Sigma}}_{K1}, ...,\widehat{\bm{\Sigma}}_{KK}) + \log(N)[(2d+1)K-1].
\end{equation}
The number of clusters used in the final analysis is then chosen to be the value $K$ that yields the lowest BIC.

\section{Simulation Studies}
\label{sec:simulations}

To evaluate the performance of the proposed elastic $k$-means approach, we simulate a series of functional samples with known ground truth cluster partitions. We assess clustering performance via the popular Rand index \citep{L_Hubert_1985}. The index is bounded above by $1$, which indicates a perfect match between generated and ground truth cluster partitions, while a value close to $0$ indicates a near-random partitioning. 

\begin{figure}[!t]
	\begin{center}
		\begin{tabular}{|c|c|}
			\hline
			$\tau=0.05$ & $\tau=0.1$\\
			\includegraphics{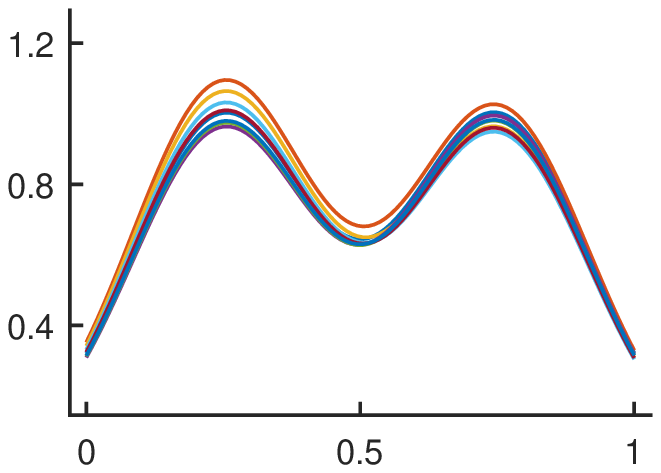}&\includegraphics{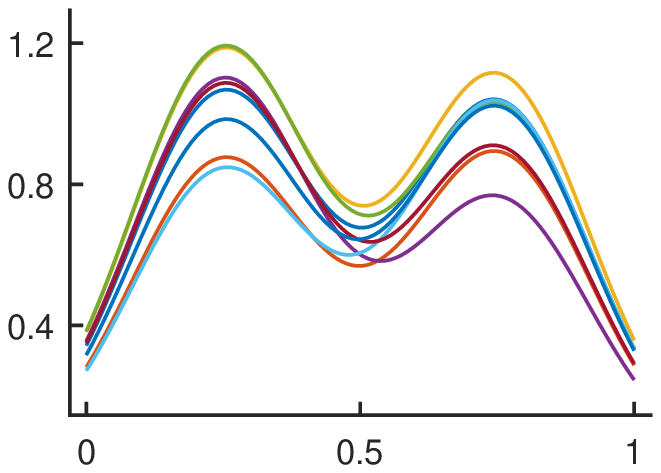}\\
			\hline
		\end{tabular}
		\caption{Generated amplitude variability for $\tau=0.05$ (left) and $\tau=0.1$ (right).}
		\label{fig:sim1taus}
	\end{center}
\end{figure}

In the first set of simulations, we consider one-dimensional functional data. We define each simulated function $g_i(t)$ as the sum of $p_i$ Gaussian kernels with random amplitude variability induced by $z_{ij} \stackrel{iid}{\sim} N(1, \tau)$: \[ g_i(t) = \sum_{j = 1}^{p_i} z_{ij}\phi\left(t;\frac{2j-1}{2p_i}, \frac{1}{3p_i}\right),\quad t \in [0, 1],  \] where $\phi(t;\mu, \sigma) = \exp\left\{-\frac{1}{2\sigma^2}\left(t-\mu\right)^2\right\}$ and $p_i = k$ with probability $\frac{1}{K^*}$ for $k=1,...,K^*$. Here, $K^*$ denotes the total number of clusters. Intuitively, when $K^*=1$, the sample only consists of single-peak functions, when $K^*=2$ the sample contains single-peak and two-peak functions, and so on. The functions $f_i$ used for clustering are generated through a random warping of $g_i$, i.e., $ f_{i} = g_i \circ \gamma_i$,
where $\gamma_i(t) = \frac{e^{\alpha_i t} - 1}{e^{\alpha_i} - 1}$ with $\alpha_i \stackrel{iid}{\sim} Uniform[-3, 3]$. We varied $K^*$ from 1 to 4, and considered two sample size settings, $N=\{120,240\}$, and two scaling variances $\tau=\{0.05,0.1\}$. The resulting within-cluster amplitude variability due to varying $\tau$ is shown in Figure \ref{fig:sim1taus}.

We compare the elastic $k$-means approach to three other approaches: (1) $k$-means with alignment (KMA) \citep{L_Sangalli_2010}, (2) standard $k$-means on the discretized functions $f_i([t])$ without alignment, and (3) standard $k$-means on discretized functions after multiple alignment $\tilde f_i([t])$; the alignment in (3) is computed using the method described in Section \ref{sec:multalign}. Note that approaches (2) and (3) essentially view the functional data after discretization as multivariate data. For KMA, we use the function kma() in the R package fdakma with default parameter settings. For elastic $k$-means, and methods (2) and (3), we repeat the clustering procedure $10$ times, with random initializations, and retain the result with the lowest value of the cost function. For KMA, we only performed the clustering once due to high computational cost. 

Table \ref{table:sim1RandIndex} reports clustering accuracy for each considered method; we report the average Rand index (with standard deviations in parentheses) across 50 replicates. In this simulation, elastic $k$-means perfectly matched the ground truth partitions for almost all of the simulation settings and replicates, outperforming the other methods tested. The accuracy of KMA was poor overall with similar performance to standard $k$-means on the unaligned, discretized functions. Interestingly, $k$-means applied to the aligned, discretized functions performed well when within-cluster amplitude variability was small, and performed poorly otherwise.

\begin{table}[!t]
	\begin{center}
		\caption{Average adjusted Rand indices (with standard deviations in parentheses), computed across 50 replicates, for (a) elastic $k$-means, (b) KMA, (c) $k$-means on unaligned, discretized functions, and (d) $k$-means on aligned, discretized functions. Best performance is highlighted in bold.}
		\label{table:sim1RandIndex}
		\begin{tabular}{*{7}{|c}|}
			\hline
			$N$ & $\tau$ & $K^*$ & (a) & (b) & (c) & (d) \\
			\hline
			\multirow{6}{*}{120} & \multirow{3}{*}{0.05} & 2 & \textbf{1.0 (0)} & 0.12 (0.17) & 0.01 (0.02) & 0.99 (0.02) \\
			& & 3 & \textbf{1.0 (0)} & 0.16 (0.11) & 0.19 (0.06) & 0.99 (0.05) \\
			& & 4 & \textbf{1.0 (0)} & 0.19 (0.11) & 0.21 (0.04) & 0.98 (0.07) \\
			\cline{2-7}
			& \multirow{3}{*}{0.1} & 2 & \textbf{1.0 (0)} & 0.09 (0.14) & 0.01 (0.02) & 0.11 (0.11) \\
			& & 3 & \textbf{1.0 (0)} & 0.17 (0.13) & 0.16 (0.05) & 0.28 (0.10) \\
			& & 4 & \textbf{1.0 (0)} & 0.15 (0.10) & 0.18 (0.05) & 0.35 (0.08) \\
			\hline
			\multirow{6}{*}{240} & \multirow{3}{*}{0.05} & 2 & \textbf{1.0 (0)} & 0.09 (0.12) & 0.011 (0.02) & 0.99 (0.01) \\
			& & 3 & \textbf{1.0 (0)} & 0.17 (0.13) & 0.18 (0.05) & 1.0 (0.01) \\
			& & 4 & \textbf{1.0 (0)} & 0.15 (0.14) & 0.20 (0.03) & 0.98 (0.07) \\
			\cline{2-7}
			& \multirow{3}{*}{0.1} & 2 & \textbf{1.0 (0)} & 0.08 (0.12) & 0.01 (0.01) & 0.08 (0.09) \\
			& & 3 & \textbf{1.0 (0)} & 0.15 (0.13) & 0.16 (0.05) & 0.27 (0.07) \\
			& & 4 & \textbf{1.0 (0.0017)} & 0.15 (0.10) & 0.18 (0.03) & 0.35 (0.06) \\
			\hline
		\end{tabular}
	\end{center}
\end{table}

\begin{figure}[!t]
	\begin{center}
		\begin{tabular}{c}			\includegraphics{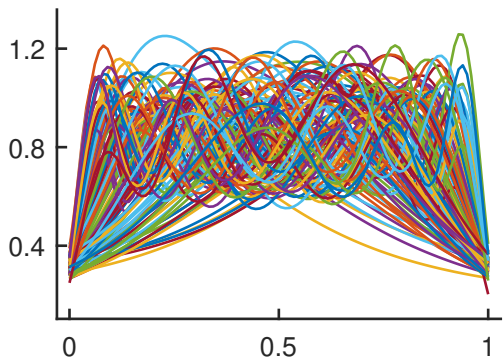}\\
		\end{tabular}
		\caption{A simulated dataset with $N = 120$, $\tau = 0.1$ and $K^*= 3$.}
		\label{fig:sim1spag}
	\end{center}
\end{figure}

\begin{figure}[!t]
	\centering
	\begin{tabular}{|c|c|c|c|}
		\hline
		 elastic-kmeans & KMA & Kmeans on $f([t])$ & Kmeans on $\tilde{f}([t])$  \\
		\includegraphics{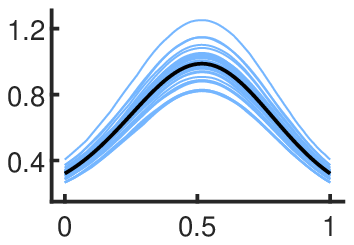} & \includegraphics{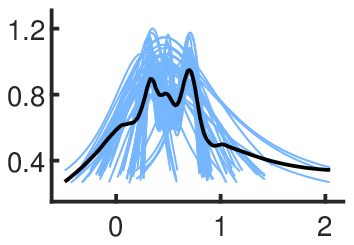} & \includegraphics{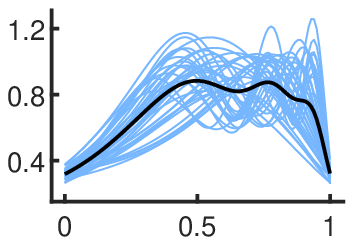} & \includegraphics{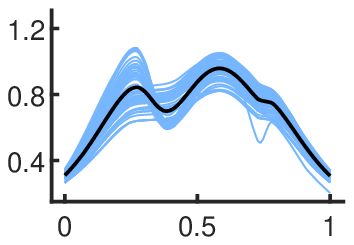} \\
		\includegraphics{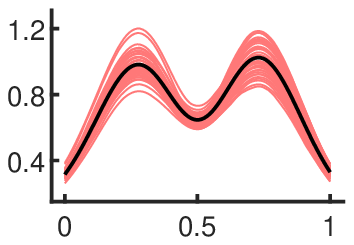} & \includegraphics{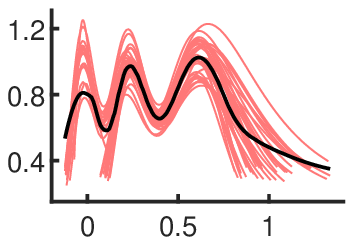} & \includegraphics{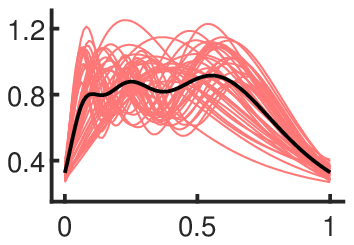} & \includegraphics{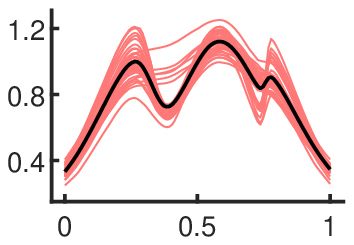} \\
		\includegraphics{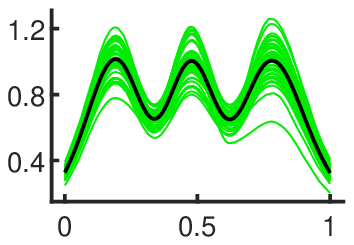} & \includegraphics{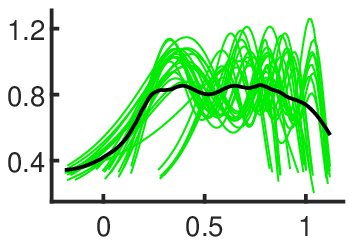} & \includegraphics{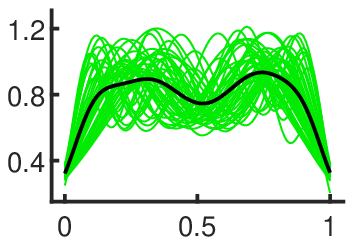} & \includegraphics{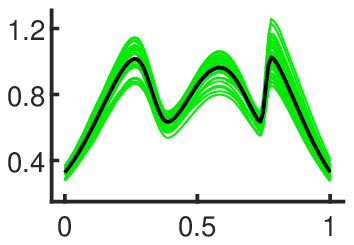} \\
		\hline
	\end{tabular}
	\caption{Clustering results for the functions shown in Figure \ref{fig:sim1spag}: (a) elastic $k$-means, (b) KMA, (c) $k$-means applied to unaligned, discretized functions, and (d) $k$-means applied to aligned, discretized functions. The template function for each cluster is shown in black.}
	\label{fig:sim1results}
\end{figure}

We looked at the results in detail for one of the simulated datasets where $N = 120$, $\tau = 0.1$ and $K^* = 3$. Figure \ref{fig:sim1spag} shows the observed functions; it is difficult to extract any meaningful information by visual inspection of the spaghetti plot. Figure \ref{fig:sim1results} displays clustering results computed using the four aforementioned methods. It is clear that the poor clustering accuracy of KMA is largely attributed to poor alignment; this is due to a lack of flexibility in warping functions considered by that method. Applying $k$-means to the unaligned, discretized functions resulted in mixed clusters, as expected. Applying $k$-means to aligned, discretized functions performed slightly better than the two already mentioned approaches. However, aligning functions with different numbers of peaks and valleys resulted in large distortions as it is unclear which features should be matched. Finally, the elastic $k$-means approach achieved perfect clustering and very good alignment of functions within each cluster.

\begin{figure}[!t]
	\centering
	\begin{tabular}{|c| c c c|}
		\hline
		\includegraphics{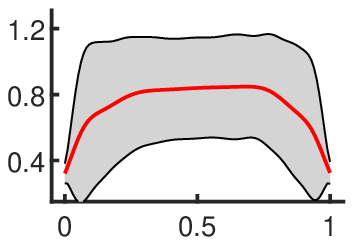} & \includegraphics{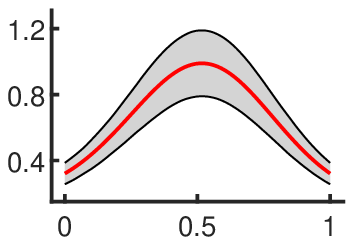} & \includegraphics{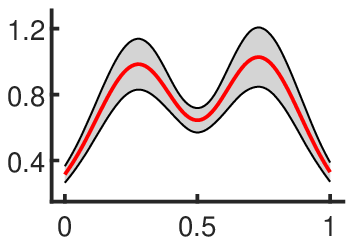} & \includegraphics{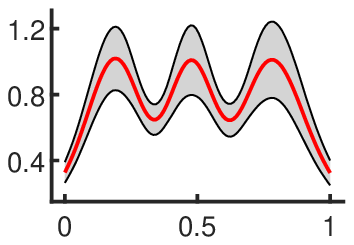} \\
		\hline
	\end{tabular}
	\caption{Pointwise summaries (mean $\pm$ 2 standard deviations) of functions in Figure \ref{fig:sim1spag} before (first panel) and after clustering and alignment via elastic $k$-means (right three panels).}
	\label{fig:sim1bands}
\end{figure}

Figure \ref{fig:sim1bands} shows that elastic $k$-means yields improved statistical summaries of functional data. The first panel depicts the cross-sectional mean $\pm 2$ standard deviations for the simulated data in Figure \ref{fig:sim1spag} before clustering and alignment. This representation falsely suggests that a typical function in this sample is mostly flat; it also exhibits very large vertical variability as evidenced by the wide error bands. On the other hand, the pointwise summaries of aligned functions within each cluster (right three panels) result in the correct three typical shapes of functions in this sample. The error bands also reveal true amplitude variability within each cluster.

\begin{figure}[!t]
	\centering
	\begin{tabular}{|c|c|c|c|c|c|}
		\hline
		$N$ & $\tau$ & $K^*=1$ & $K^*=2$ & $K^*=3$ & $K^*=4$\\
		\hline
		\multirow{2}{*}{120} & 0.05 & \includegraphics[width=\simonebicwid]{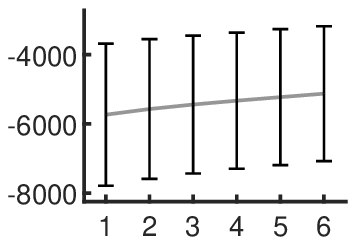} & \includegraphics[width=\simonebicwid]{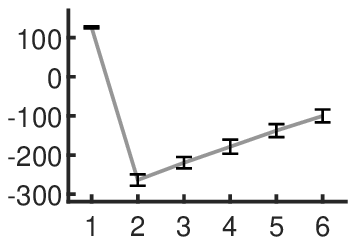} &
		\includegraphics[width=\simonebicwid]{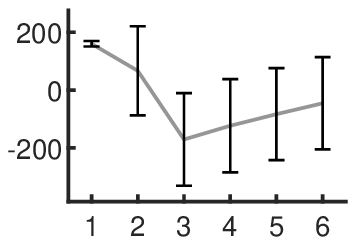} &
		\includegraphics[width=\simonebicwid]{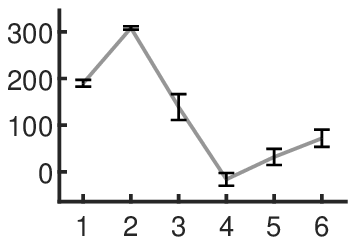} \\
		\cline{2-6}
		& 0.1 & 
		\includegraphics[width=\simonebicwid]{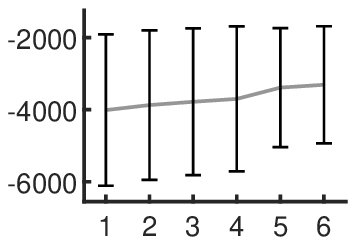} &
		\includegraphics[width=\simonebicwid]{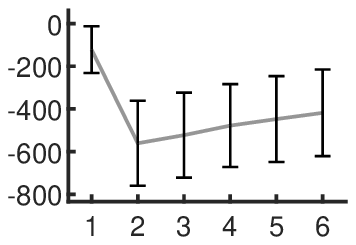} &
		\includegraphics[width=\simonebicwid]{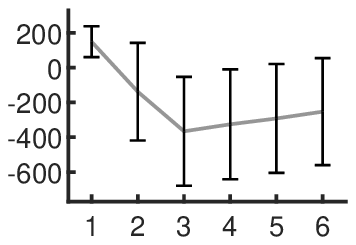} &
		\includegraphics[width=\simonebicwid]{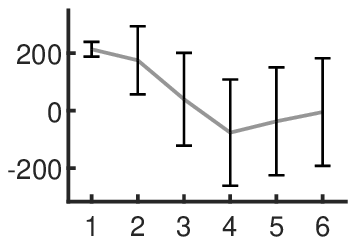} \\
		\hline
		\multirow{2}{*}{240} & 0.05 & \includegraphics[width=\simonebicwid]{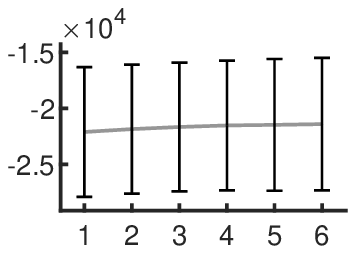} & \includegraphics[width=\simonebicwid]{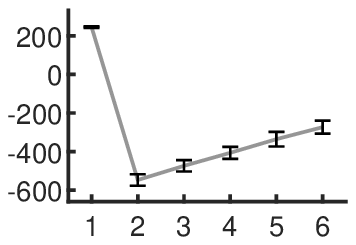} &
		\includegraphics[width=\simonebicwid]{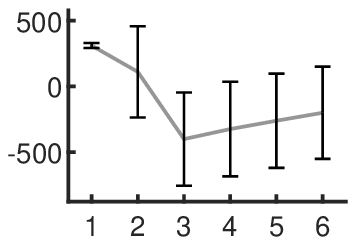} &
		\includegraphics[width=\simonebicwid]{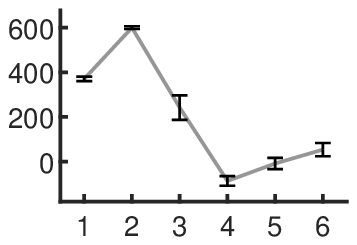} \\
		\cline{2-6}
		& 0.1 & 
		\includegraphics[width=\simonebicwid]{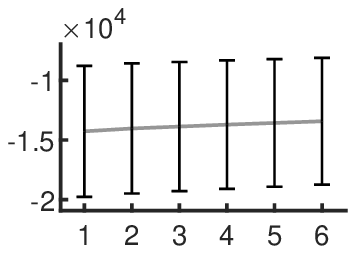} &
		\includegraphics[width=\simonebicwid]{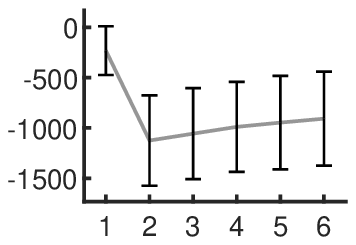} &
		\includegraphics[width=\simonebicwid]{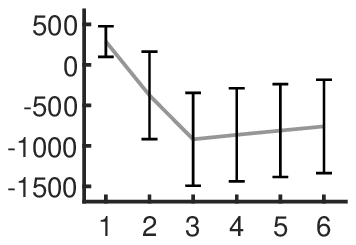} &
		\includegraphics[width=\simonebicwid]{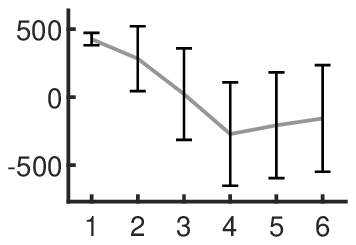} \\
		\hline
	\end{tabular}
	\caption{Selection of $K$ in Simulation 1 using the proposed BIC criterion. Error bars show one standard deviation around the average BIC across 50 replicates.}
	\label{fig:sim1bic}
\end{figure}

The results reported so far were computed under the assumption that the true number of clusters $K^*$ is known. However, in practice, the number of clusters should also be inferred based on the given data. For the elastic $k$-means approach, we use the proposed BIC criterion to select the number of clusters. Thus, we next assess the effectiveness of this criterion; to reduce dimension when computing the BIC, we set $\rho=0.95$. The minimum and maximum numbers of clusters allowed are set to $1$ and $6$, respectively. The data-generating process in this simulation is the same as in the previous one. Figure \ref{fig:sim1bic} shows the average BIC for different true values of $K^*$ across 50 replicates (with standard deviations shown as error bars). When the true number of clusters $K^*>1$, the proposed criterion always chooses the correct number of clusters. The corresponding error bars are also narrow, providing further confidence in the results. 

The task becomes more challenging when $K^* = 1$, i.e., there is no clustering structure in the data. In this case, the data is dominated by random within-cluster variability and no systematic between-cluster variability. As a result, the variation in computed BIC values is much larger in this case. Nevertheless, in the vast majority of replicates, the proposed BIC criterion chose $1$ as the number of clusters. In fact, the wrong number of clusters was indicated in only two out of the total 200 replicates (50 replicates for each choice of $N$ and $\tau$). 

The proposed approach is also applicable to clustering of vector-valued functional data. To illustrate its performance relative to existing approaches, a similar simulation study was conducted for two-dimensional functional data as described in Supplementary Material Section \ref{supsec1}. 
Another simulation study, which is more closely linked to our motivating application, is presented in Supplementary Material Section \ref{supsec2} and relates to posterior visualization of dynamical system states. We applied elastic $k$-means to the problem of clustering posterior sample states of the FitzHugh-Nagumo system of ordinary differential equations. Although in this case a ground truth clustering is not available, the performance of the proposed approach was assessed visually relative to clustering obtained for a model parameter that controls phase. 

\section{Clustering for Posterior Visualization of ARI Trajectories}
\label{sec:application}

\begin{figure}[!t]
	\begin{center}
		\begin{tabular}{|c|c|c|}
			\hline
			2002-03 & 2003-04 & 2004-05 \\
			\includegraphics{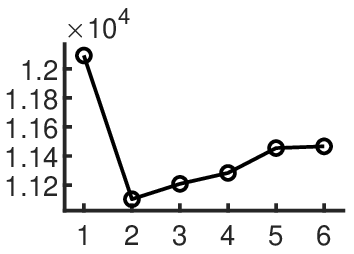} &
			\includegraphics{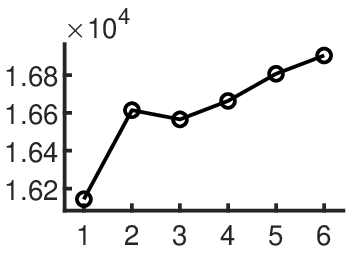} &
			\includegraphics{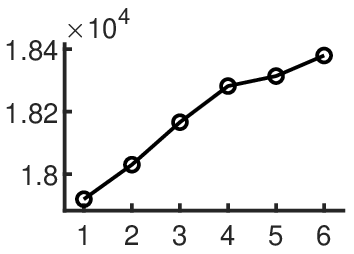} \\
			\hline
			2005-06 & 2006-07 & 2007-08\\
			\includegraphics{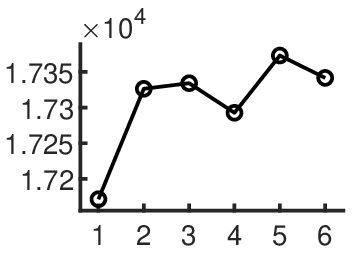} &
			\includegraphics{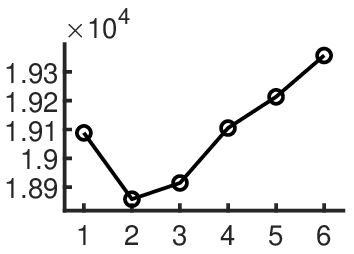} &
			\includegraphics{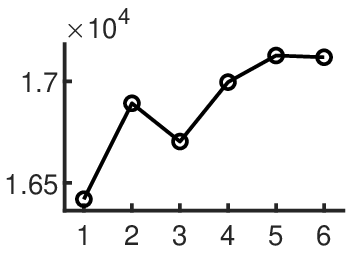} \\
			\hline
		\end{tabular}
		\caption{The number of clusters, $K$, for the posterior aggregated ARI trajectories in each epidemic year was selected by applying the proposed BIC criterion to 400 posterior samples.}
		\label{fig:ari_bic}
	\end{center}
\end{figure}

\begin{figure}[!t]
	\begin{center}
		\begin{tabular}{|c|c c|}
			\hline
			& (a) & (b) \\
			\hline
			(1) &
			
			\includegraphics{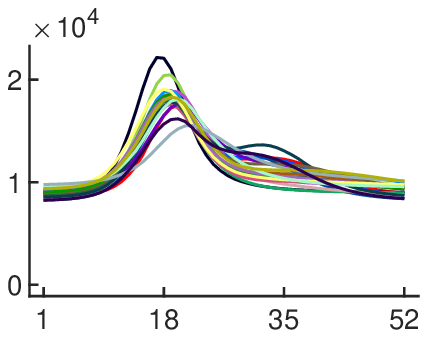} &
			\includegraphics{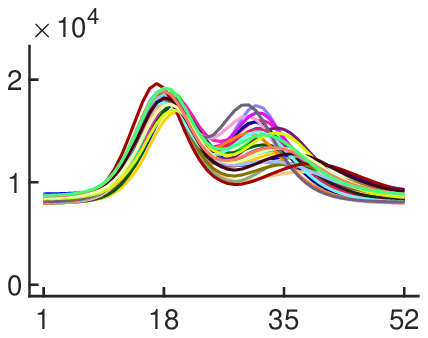} \\
			\hline
			(2) &
			
			\includegraphics{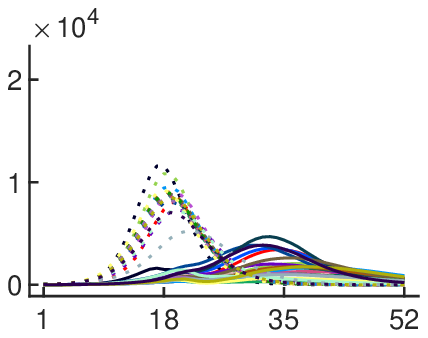} &
			\includegraphics{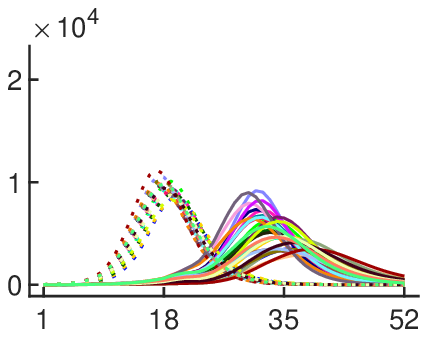} \\
			\hline
		\end{tabular}
		\caption{Elastic $k$-means clustering based on 400 posterior draws of aggregated posterior ARI trajectories for the infection year 2002-03. Row (1): Unaligned functions in each cluster. Row (2): Disaggregated influenza (solid line) and RSV (dotted line) functions in each cluster. Columns (a) and (b) correspond to the two clusters shown in Figure \ref{fig:ari_2002}. Display is based on a down-sampling of 50 functions for clarity.}
		\label{fig:ari_additional_2002_50}
	\end{center}
\end{figure}

We applied our method to the problem of clustering posterior aggregated ARI trajectories in each of the six epidemic years from 2002-03 to 2008-09. The BIC criterion suggests that there are two amplitude clusters in the epidemic years 2002-03 and 2006-07, and that there is only a single cluster in all other years (Figure \ref{fig:ari_bic}). Of the two clusters identified in 2002-03, one contains a single infection peak, while the other shows a first major peak followed by a smaller second peak (Figures \ref{fig:ari_2002}, \ref{fig:ari_additional_2002_50}).  \citet{Y_Garcia_2017} jointly modelled the interaction of influenza and RSV infection states, resulting in posterior infection trajectories that could be disaggregated by pathogen. We therefore inspected these disaggregated trajectories in each cluster (Figure \ref{fig:ari_additional_2002_50}) and found that the initial major ARI outbreak in both clusters could be attributed to a peak of RSV infection early in the epidemic year. The minor second ARI peak in the two-outbreak cluster was due to either a small influenza infection peak, that occurred much later in the epidemic year and is well-separated from the RSV infection peak, or an influenza peak that is relatively close to the RSV peak in terms of timing, but is larger in magnitude. Importantly, this two-cluster structure could be identified neither visually from the spaghetti plot, nor from the pointwise summaries of functions without clustering and alignment (lower panel of Figure \ref{fig:ari_2002}(a)). Furthermore, the pointwise upper bound, mean, and lower bound of the original functions attain their peaks at different times. However, the distortion of the summaries around the second peak is more evident, which suggests that phase variability is distorting the summaries. Another interesting observation is that the pointwise error band forms a bulge around the second infection peak, but clustering and alignment allow us to attribute this behaviour to both the existence of multiple clusters of trajectories in the sample and the increase of variability in the second outbreak of the two-peak cluster. 

Clustering of aggregated posterior ARI samples from 2006-07, shown in Supplementary Figure \ref{fig:ari_2006}, did not provide as much additional insight into the two-pathogen interaction as in 2002-03, but still suggested a possible heterogeneity in the posterior, mainly due to the height of the infection peak. The posterior trajectories for infection years in which BIC suggested the absence of multiple clusters are shown in Supplementary Figure \ref{fig:years_wo_clusters}.

\begin{figure}[!t]
	\begin{center}
		\begin{tabular}{|c|c|c|}
			\hline
			2002-03 & 2003-04 & 2004-05 \\
			\includegraphics{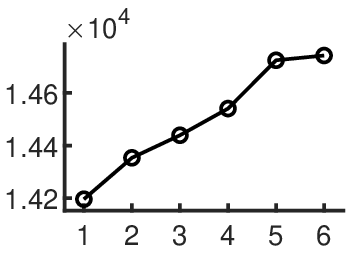} &
			\includegraphics{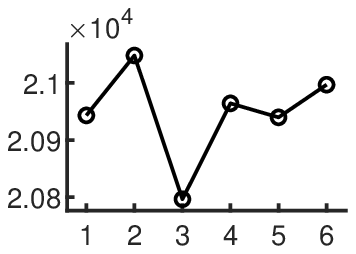} &
			\includegraphics{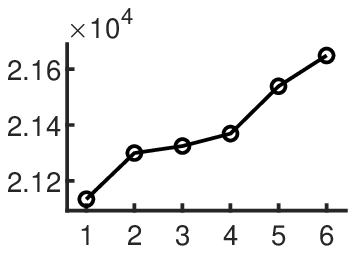} \\
			\hline
			2005-06 & 2006-07 & 2007-08\\
			\includegraphics{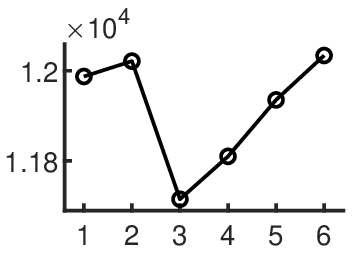} &
			\includegraphics{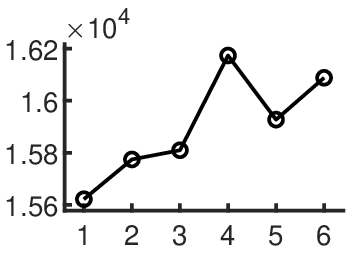} &
			\includegraphics{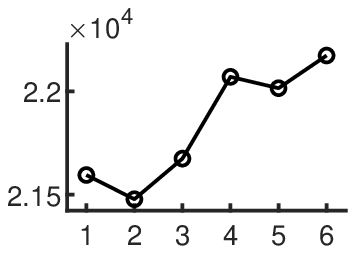} \\
			\hline
		\end{tabular}
		\caption{Number of clusters, $K$, selected via the proposed BIC for the disaggregated influenza and RSV posterior trajectories, based on 400 posterior samples.}
		\label{fig:ari_2d_bic}
	\end{center}
\end{figure}

\begin{figure}[!t]
	\begin{center}
		\begin{tabular}{|c|c|c|}
		    \hline
			\includegraphics{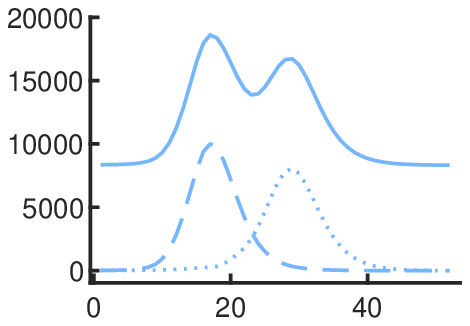} &
			\includegraphics{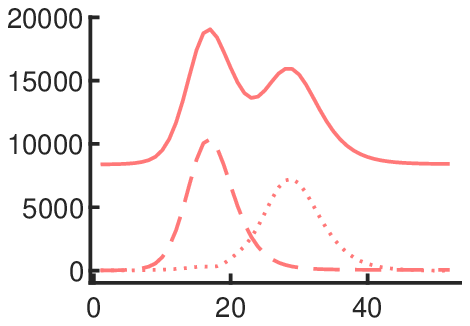} &
			\includegraphics{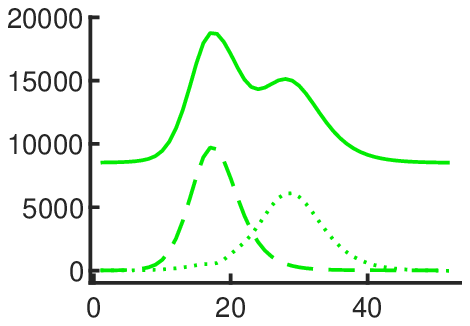} \\
			\hline
		\end{tabular}
		\caption{Elastic $k$-means clustering based on 400 posterior draws of disaggregated influenza and RSV infection trajectories for the infection year 2003-04. For clarity, we only show the template function for each cluster. Dashed line: Influenza. Dotted line: RSV. Solid line: Aggregated template with average background infections added. $K = 3$ is identified by the BIC criterion.}
		\label{fig:ari_2d_2003}
	\end{center}
\end{figure}

\begin{figure}[!t]
	\begin{center}
		\begin{tabular}{|c|c|c|}
		    \hline
			\includegraphics{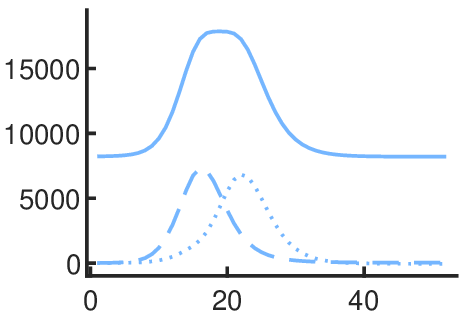} &
			\includegraphics{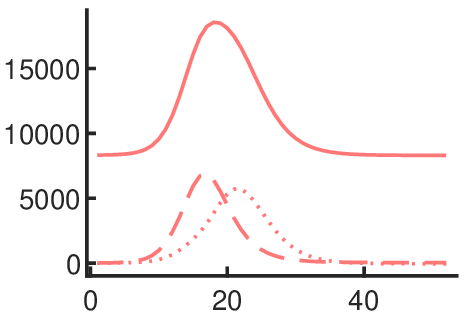} &
			\includegraphics{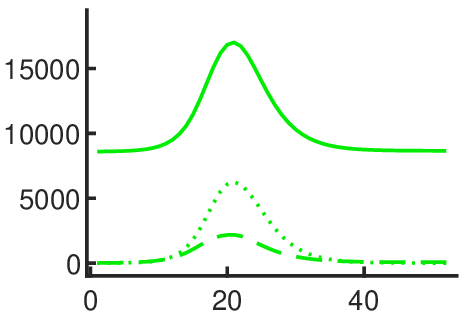} \\
			\hline
		\end{tabular}
		\caption{Elastic $k$-means clustering based on 400 posterior draws of disaggregated influenza and RSV infection trajectories for the infection year 2005-06. For clarity, we only show the template function for each cluster. Dashed line: Influenza. Dotted line: RSV. Solid line: Aggregated template with average background infections added. $K = 3$ is identified by the BIC criterion.}
		\label{fig:ari_2d_2005}
	\end{center}
\end{figure}

\begin{figure}[!t]
	\begin{center}
		\begin{tabular}{|c|c|}
		    \hline
			\includegraphics{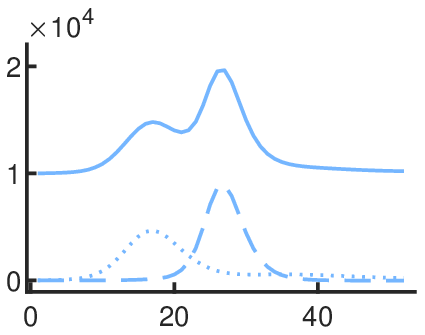} &
			\includegraphics{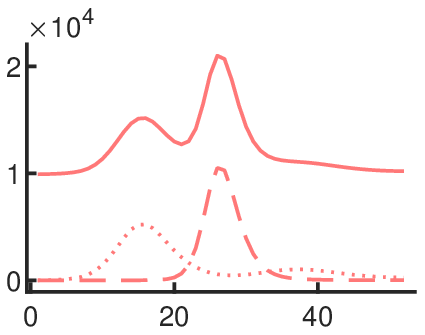} \\
			\hline
		\end{tabular}
		\caption{Elastic $k$-means clustering based on 400 posterior draws of disaggregated influenza and RSV infection trajectories for the infection year 2007-08. For clarity, we only show the template function for each cluster. Dashed line: Influenza. Dotted line: RSV. Solid line: Aggregated template with average background infections added. $K = 2$ is identified by the BIC criterion.}
		\label{fig:ari_2d_2007}
	\end{center}
\end{figure}

Since posterior sample trajectories over influenza and RSV infections are available separately, an alternative way to apply our approach to the exploration of ARI posterior structure is to jointly view the posterior samples as vector-valued functions $f: [1,52] \rightarrow \mathbb{R}^2$, where the first dimension corresponds to influenza and the second to RSV. The BIC method suggests three clusters for year 2003-04 and 2005-06, two clusters for 2007-08 and no clustering structure for other years (Figure \ref{fig:ari_2d_bic}). We notice that in most years selection of $K$ in disaggregated influenza-ARI vector-valued functions leads to conclusions different from those given by selection of $K$ in aggregated ARI trajectories (Figure \ref{fig:ari_bic}). This is not surprising, as aggregating infection trajectories from different pathogens in a pointwise fashion leads to loss of information. For infection years 2003-04 and 2005-06 where no clustering structure was identified in aggregated ARI trajectories, elastic $k$-means clustering of disaggregated influenza and RSV infection trajectories identified clusters with different relative magnitudes of influenza versus RSV infections, and different extent of overlap between them (Figures \ref{fig:ari_2d_2003}, \ref{fig:ari_2d_2007} and Supplementary Figures \ref{fig:supp_ari2d_2003_50}, \ref{fig:supp_ari2d_2007_50}). In infection year 2005-06, the aggregated infections consist of a single cluster with one infection peak, while the disaggregated samples yield three clusters. In two of these clusters, influenza and RSV peaks occur at slightly different times with large, but partial, overlap and with different relative magnitudes. In the remaining cluster, the peak infections are completely synchronized with the RSV peak being much higher than the influenza peak (Figure \ref{fig:ari_2d_2005} and Supplementary Figure \ref{fig:supp_ari2d_2005_50}). On the other hand, for infection year 2002-03, only a single cluster was found in disaggregated influenza-RSV trajectories  (Supplementary Figures \ref{fig:ari_2d_1cl}, \ref{fig:2d_years_wo_clusters}), which seems to be discordant with the two distinct overall ARI clusters shown in Figures \ref{fig:ari_2002} and \ref{fig:ari_additional_2002_50}. The lack of clustering structure in disaggregated functions is largely due to the continuous spectrum of difference in influenza versus RSV infection peaks. Aggregation of those two dimensions, however, leads to some aggregated ARI trajectories having two distinct peaks with others having a single peak. 


\section{Discussion}
\label{sec:discussion}

The elastic functional data analysis framework has been shown to be effective for the problems of functional alignment and shape registration \citep{A_Srivastava_2016}. In this paper, we showed that adoption of the same framework towards functional $k$-means greatly enhances the accuracy of clustering results compared to KMA, which shares the same algorithmic design, by virtue of the flexible group of warping functions that leads to better alignment quality within clusters. We additionally proposed a model-based approach to choose the optimal number of clusters by minimizing a BIC computed via cluster-wise fPCA. When it works together with elastic $k$-means and pointwise summaries applied to aligned functions within each cluster, they form a pipeline producing more informative summaries of Bayesian posterior sample functions. 

Since elastic $k$-means effectively performs multiple alignments in each cluster, the within-cluster amplitude variability is encoded by the aligned functions, while within-cluster phase variability is encoded by the optimal warpings. In the results, we only showed the visualization of the amplitude variability, but one can certainly visualize the phase variability by applying similar pointwise summaries to the warping functions. Our method resolves the distortion of shape and exaggeration of amplitude variability in the presence of misalignment and multiple clusters, but an additional criticism over the pointwise summaries is that they ignore the underlying functional structure, which is resolved by taking the perspective of functional data analysis. Alternative visualizations of within-cluster amplitude and phase variability that preserves the functional structure could be considered for further improvement, such as the functional boxplot-type display devised by \citet{X_Weiyi_2017}. 

We mentioned that elastic $k$-means is an amplitude clustering method that treats phase variability as irrelevant to clustering, as could be seen from the cost function that depends on amplitude distance, but not on phase distance. Therefore, it may not be suitable for some application scenarios, especially when phase plays an important role in the clustering structure of interest. For example, in the Berkeley Growth Study \citep{H_Jones_1941}, one would expect that phase is a pivotal factor distinguishing growth velocity curves of two genders, as growth spurts occur at different time points in males and females. If we use just the amplitude component for clustering, it will not be surprising to see the cluster labels matching poorly with genders. 
In the future, we plan to develop a clustering approach that incorporates the phase distance into the cost function, and gives users the freedom to control the importance weight of amplitude versus phase.

When the functions are vector-valued, an implicit assumption of elastic $k$-means is that phase is synchronized across multiple dimensions, as we always apply the same warping to each dimension. Thus, another interesting research direction is to allow different warpings across dimensions, possibly with a certain penalty imposed on the asynchrony.

\section*{Acknowledgments}

The authors gratefully acknowledge Drs. Yury Garc\'ia, Marcos Capistr\'an (Centro de Investigaci\'on en Matem\'aticas, Guanajuato, Gto., M\'exico), and Daniel Noyola (Department of Microbiology, Faculty of Medicine, Universidad Aut\'onoma de San Luis Potos\'i) for the use of posterior samples from their analysis of acute respiratory infections in San Luis Potos\'i, M\'exico. 

This paper describes objective technical results and analysis. Any subjective views or opinions that might be expressed in the paper do not necessarily represent the views of the U.S. Department of Energy or the United States Government. 
This work was supported by the Laboratory Directed Research and Development program at Sandia National Laboratories, a multi-mission laboratory managed and operated by National Technology and Engineering Solutions of Sandia,
LLC, a wholly owned subsidiary of Honeywell International, Inc., for the
U.S. Department of Energy's National Nuclear Security Administration
under contract DE-NA0003525. This research was partially funded by NSF DMS-1613054, NSF CCF-1740761, NSF DMS-2015226, NSF CCF-1839252 and NIH R37-CA214955. 

\clearpage
\bibliographystyle{chicago}
\bibliography{FRkmeansJournal}

\begin{thebibliography}{}

\bibitem[\protect\citeauthoryear{Abraham, Cornillon, Matzner-Løber, and
  Molinari}{Abraham et~al.}{2003}]{C_Abraham_2003}
Abraham, C., P.~A. Cornillon, E.~Matzner-Løber, and N.~Molinari (2003).
\newblock Unsupervised curve clustering using {B}-splines.
\newblock {\em Scandinavian Journal of Statistics\/}~{\em 30\/}(3), 581--595.

\bibitem[\protect\citeauthoryear{Allen}{Allen}{2017}]{L_Allen_2017}
Allen, L. J.~S. (2017).
\newblock A primer on stochastic epidemic models: Formulation, numerical
  simulation, and analysis.
\newblock {\em Infectious Disease Modelling\/}~{\em 2\/}(2), 128--142.

\bibitem[\protect\citeauthoryear{Baetica, Catanach, Hsiao, Murray, and
  Beck}{Baetica et~al.}{2016}]{A_Baetica_2016}
Baetica, A.~A., T.~A. Catanach, V.~Hsiao, R.~M. Murray, and J.~L. Beck (2016).
\newblock A {B}ayesian approach to inferring chemical signal timing and
  amplitude in a temporal logic gate using the cell population distributional
  response.
\newblock {\em bioRxiv\/}, 087379.

\bibitem[\protect\citeauthoryear{Banfield and Raftery}{Banfield and
  Raftery}{1993}]{J_Banfield_1993}
Banfield, J.~D. and A.~E. Raftery (1993).
\newblock Model-based {G}aussian and non-{G}aussian clustering.
\newblock {\em Biometrics\/}~{\em 49\/}(3), 803--821.

\bibitem[\protect\citeauthoryear{Bradley, Bennett, and Demiriz}{Bradley
  et~al.}{2000}]{P_Bradley_2000}
Bradley, P.~S., K.~P. Bennett, and A.~Demiriz (2000).
\newblock Constrained k-means clustering.
\newblock {\em Microsoft Research, Redmond\/}~{\em 20}.

\bibitem[\protect\citeauthoryear{Eisenkolb, Jensch, Eisenkolb, Kramer,
  Buchholz, Pleiss, Spiess, and Radde}{Eisenkolb
  et~al.}{2019}]{I_Eisenkolb_2019}
Eisenkolb, I., A.~Jensch, K.~Eisenkolb, A.~Kramer, P.~C.~F. Buchholz,
  J.~Pleiss, A.~Spiess, and N.~E. Radde (2019).
\newblock Modeling of biocatalytic reactions: A workflow for model calibration,
  selection and validation using {B}ayesian statistics.
\newblock {\em AIChE Journal\/}, e16866.

\bibitem[\protect\citeauthoryear{Forgy}{Forgy}{1965}]{E_Forgy_1965}
Forgy, E.~W. (1965).
\newblock Cluster analysis of multivariate data: Efficiency versus
  interpretability of classifications.
\newblock {\em Biometrics\/}~{\em 21}, 768--769.

\bibitem[\protect\citeauthoryear{García, Chkrebtii, Capistrán, and
  Noyola}{García et~al.}{2017}]{Y_Garcia_2017}
García, Y.~E., O.~A. Chkrebtii, M.~A. Capistrán, and D.~E. Noyola (2017).
\newblock Identifying individual disease dynamics in a stochastic
  multi-pathogen model from aggregated reports and laboratory data.
\newblock {\em arXiv\/}, 1710.10346v1.

\bibitem[\protect\citeauthoryear{Gelman, Carlin, Stern, Dunson, Vehtari, and
  Rubin}{Gelman et~al.}{2013}]{A_Gelman_2013}
Gelman, A., J.~B. Carlin, H.~S. Stern, D.~B. Dunson, A.~Vehtari, and D.~B.
  Rubin (2013).
\newblock {\em {B}ayesian data analysis}.
\newblock Chapman and Hall/CRC.

\bibitem[\protect\citeauthoryear{Grinstein, Trutschl, and Cvek}{Grinstein
  et~al.}{2001}]{G_Grinstein_2001}
Grinstein, G., M.~Trutschl, and U.~Cvek (2001).
\newblock High-dimensional visualizations.
\newblock In {\em Proceedings of the Visual Data Mining Workshop}, Volume~2.

\bibitem[\protect\citeauthoryear{Huang, Gallivan, Srivastava, and Absil}{Huang
  et~al.}{2016}]{huang2016}
Huang, W., K.~A. Gallivan, A.~Srivastava, and P.~Absil (2016).
\newblock Riemannian optimization for registration of curves in elastic shape
  analysis.
\newblock {\em Journal of Mathematical Imaging and Vision\/}~{\em 54\/}(3),
  320--343.

\bibitem[\protect\citeauthoryear{Hubert and Arabie}{Hubert and
  Arabie}{1985}]{L_Hubert_1985}
Hubert, L. and P.~Arabie (1985).
\newblock Comparing partitions.
\newblock {\em Journal of Classification\/}~{\em 2\/}(1), 193--218.

\bibitem[\protect\citeauthoryear{Hyndman and Shang}{Hyndman and
  Shang}{2010}]{R_Hyndman_2010}
Hyndman, R.~J. and H.~L. Shang (2010).
\newblock Rainbow plots, bagplots, and boxplots for functional data.
\newblock {\em Journal of Computational and Graphical Statistics\/}~{\em
  19\/}(1), 29--45.

\bibitem[\protect\citeauthoryear{Iglesiasm, Lin, and Stuart}{Iglesiasm
  et~al.}{2014}]{IglesiasmEtAl2014}
Iglesiasm, M., K.~Lin, and A.~Stuart (2014).
\newblock Well-posed {B}ayesian geometric inverse problems arising in
  subsurface flow.
\newblock {\em Inverse Problems\/}~{\em 30\/}(11).

\bibitem[\protect\citeauthoryear{Jacques and Preda}{Jacques and
  Preda}{2014}]{J_Jacques_2014}
Jacques, J. and C.~Preda (2014).
\newblock Functional data clustering: A survey.
\newblock {\em Advances in Data Analysis and Classification\/}~{\em 8\/}(3),
  231--255.

\bibitem[\protect\citeauthoryear{James and Sugar}{James and
  Sugar}{2003}]{G_James_2003}
James, G.~M. and C.~A. Sugar (2003).
\newblock Clustering for sparsely sampled functional data.
\newblock {\em Journal of the American Statistical Association\/}~{\em
  98\/}(462), 397--408.

\bibitem[\protect\citeauthoryear{Jones and Bayley}{Jones and
  Bayley}{1941}]{H_Jones_1941}
Jones, H.~E. and N.~Bayley (1941).
\newblock The {B}erkeley growth study.
\newblock {\em Child Development\/}~{\em 12\/}(2), 167--173.

\bibitem[\protect\citeauthoryear{Kruschke}{Kruschke}{2014}]{J_Kruschke_2014}
Kruschke, J. (2014).
\newblock {\em Doing {B}ayesian data analysis: A tutorial with R, JAGS, and
  Stan}.
\newblock Academic Press.

\bibitem[\protect\citeauthoryear{Liu, Maljovec, Wang, Bremer, and Pascucci}{Liu
  et~al.}{2017}]{S_Liu_2017}
Liu, S., D.~Maljovec, B.~Wang, P.~Bremer, and V.~Pascucci (2017).
\newblock Visualizing high-dimensional data: Advances in the past decade.
\newblock {\em IEEE Transactions on Visualization and Computer Graphics\/}~{\em
  23\/}(3), 1249--1268.

\bibitem[\protect\citeauthoryear{Liu and Müller}{Liu and
  Müller}{2004}]{X_Liu_2004}
Liu, X. and H.-G. Müller (2004).
\newblock Functional convex averaging and synchronization for time-warped
  random curves.
\newblock {\em Journal of the American Statistical Association\/}~{\em
  99\/}(467), 687--699.

\bibitem[\protect\citeauthoryear{Liu and Yang}{Liu and Yang}{2009}]{X_Liu_2009}
Liu, X. and M.~C.~K. Yang (2009).
\newblock Simultaneous curve registration and clustering for functional data.
\newblock {\em Computational Statistics and Data Analysis\/}~{\em 53\/}(4),
  1361--1376.

\bibitem[\protect\citeauthoryear{López-Pintado and Romo}{López-Pintado and
  Romo}{2009}]{S_Lopez_2009}
López-Pintado, S. and J.~Romo (2009).
\newblock On the concept of depth for functional data.
\newblock {\em Journal of the American Statistical Association\/}~{\em
  104\/}(486), 718--734.

\bibitem[\protect\citeauthoryear{Marron, Ramsay, Sangalli, and
  Srivastava}{Marron et~al.}{2015}]{J_Marron_2015}
Marron, J.~S., J.~O. Ramsay, L.~M. Sangalli, and A.~Srivastava (2015).
\newblock Functional data analysis of amplitude and phase variation.
\newblock {\em Statistical Science\/}~{\em 30\/}(4), 468--484.

\bibitem[\protect\citeauthoryear{McDermott, Wikle, and Millspaugh}{McDermott
  et~al.}{2017}]{McDermottEtAl2017}
McDermott, P.~L., C.~K. Wikle, and J.~Millspaugh (2017).
\newblock Hierarchical nonlinear spatio-temporal agent-based models for
  collective animal movement.
\newblock {\em Journal of Agricultural, Biological and Environmental
  Statistics\/}~{\em 22\/}(3), 294--312.

\bibitem[\protect\citeauthoryear{Mio, Srivastava, and Joshi}{Mio
  et~al.}{2007}]{W_Mio_2007}
Mio, W., A.~Srivastava, and S.~Joshi (2007).
\newblock On shape of plane elastic curves.
\newblock {\em International Journal of Computer Vision\/}~{\em 73\/}(3),
  307--324.

\bibitem[\protect\citeauthoryear{Ramsay and Li}{Ramsay and
  Li}{1998}]{JO_Ramsay_1998}
Ramsay, J.~O. and X.~Li (1998).
\newblock Curve registration.
\newblock {\em Journal of the Royal Statistical Society: Series B\/}~{\em
  60\/}(2), 351--363.

\bibitem[\protect\citeauthoryear{Rao}{Rao}{1945}]{C_Rao_1945}
Rao, C.~R. (1945).
\newblock Information and the accuracy attainable in the estimation of
  statistical parameters.
\newblock {\em Resonance - Journal of Science Education\/}~{\em 20}, 78--90.

\bibitem[\protect\citeauthoryear{Rasmussen, Volz, and Koelle}{Rasmussen
  et~al.}{2014}]{D_Rasmussen_2014}
Rasmussen, D.~A., E.~M. Volz, and K.~Koelle (2014).
\newblock Phylodynamic inference for structured epidemiological models.
\newblock {\em PLOS Computational Biology\/}~{\em 10\/}(4), e1003570.

\bibitem[\protect\citeauthoryear{Robinson}{Robinson}{2012}]{D_Robinson_2012}
Robinson, D.~T. (2012).
\newblock Functional data analysis and partial shape matching in the square
  root velocity framework.
\newblock {\em PhD Thesis, Florida State University\/}.

\bibitem[\protect\citeauthoryear{Rosales, Fill, and Escobar}{Rosales
  et~al.}{2004}]{RA_Rosales_2004}
Rosales, R.~A., M.~Fill, and A.~L. Escobar (2004).
\newblock Calcium regulation of single ryanodine receptor channel gating
  analyzed using {HMM/MCMC} statistical methods.
\newblock {\em Journal of General Physiology\/}~{\em 123\/}(5), 533--53.

\bibitem[\protect\citeauthoryear{Sangalli, Secchi, Vantini, and
  Vitelli}{Sangalli et~al.}{2010}]{L_Sangalli_2010}
Sangalli, L.~M., P.~Secchi, S.~Vantini, and V.~Vitelli (2010).
\newblock K-mean alignment for curve clustering.
\newblock {\em Computational Statistics and Data Analysis\/}~{\em 54\/}(5),
  1219--1233.

\bibitem[\protect\citeauthoryear{Saxena, Prasad, Gupta, Bharill, Patel, Tiwari,
  Er, Ding, and Lin}{Saxena et~al.}{2017}]{A_Saxena_2017}
Saxena, A., M.~Prasad, A.~Gupta, N.~Bharill, O.~P. Patel, A.~Tiwari, M.~J. Er,
  W.~Ding, and C.-T. Lin (2017).
\newblock A review of clustering techniques and developments.
\newblock {\em Neurocomputing\/}~{\em 267}, 664--681.

\bibitem[\protect\citeauthoryear{Srivastava, Klassen, Joshi, and
  Jermyn}{Srivastava et~al.}{2011}]{A_Srivastava_2011_b}
Srivastava, A., E.~Klassen, S.~H. Joshi, and I.~H. Jermyn (2011).
\newblock Shape analysis of elastic curves in euclidean spaces.
\newblock {\em IEEE Transactions on Pattern Analysis and Machine
  Intelligence\/}~{\em 33\/}(7), 1415--1428.

\bibitem[\protect\citeauthoryear{Srivastava and Klassen}{Srivastava and
  Klassen}{2016}]{A_Srivastava_2016}
Srivastava, A. and E.~P. Klassen (2016).
\newblock {\em Functional and shape data analysis}.
\newblock Springer.

\bibitem[\protect\citeauthoryear{Srivastava, Wu, Kurtek, Klassen, and
  Marron}{Srivastava et~al.}{2011}]{A_Srivastava_2011}
Srivastava, A., W.~Wu, S.~Kurtek, E.~Klassen, and J.~S. Marron (2011).
\newblock Registration of functional data using {F}isher-{R}ao metric.
\newblock {\em arXiv\/}, 1103.3817.

\bibitem[\protect\citeauthoryear{Sun and Genton}{Sun and
  Genton}{2011}]{Y_Sun_2011}
Sun, Y. and M.~G. Genton (2011).
\newblock Functional boxplots.
\newblock {\em Journal of Computational and Graphical Statistics\/}~{\em
  20\/}(2), 316--334.

\bibitem[\protect\citeauthoryear{Tarpey and Kinateder}{Tarpey and
  Kinateder}{2003}]{T_Tarpey_2003}
Tarpey, T. and K.~K.~J. Kinateder (2003).
\newblock Clustering functional data.
\newblock {\em Journal of Classification\/}~{\em 20\/}(1), 093--114.

\bibitem[\protect\citeauthoryear{Tucker, Wu, and Srivastava}{Tucker
  et~al.}{2013}]{tucker2013}
Tucker, J.~D., W.~Wu, and A.~Srivastava (2013).
\newblock Generative models for functional data using phase and amplitude
  separation.
\newblock {\em Computational Statistics and Data Analysis\/}~{\em 61}, 50--66.

\bibitem[\protect\citeauthoryear{Venna and Kaski}{Venna and
  Kaski}{2003}]{J_Venna_2003}
Venna, J. and S.~Kaski (2003).
\newblock Visualizing high-dimensional posterior distributions in {B}ayesian
  modeling.
\newblock In {\em Proceedings of Artificial Neural Networks and Neural
  Information Processing}, pp.\  165--168.

\bibitem[\protect\citeauthoryear{Ward}{Ward}{1963}]{J_Ward_1963}
Ward, J.~H. (1963).
\newblock Hierarchical grouping to optimize an objective function.
\newblock {\em Journal of the American Statistical Association\/}~{\em
  58\/}(301), 236--244.

\bibitem[\protect\citeauthoryear{Wilkinson}{Wilkinson}{2011}]{Wilkinson2011}
Wilkinson, D. (2011).
\newblock {\em {S}tochastic {M}odelling for {S}ystems {B}iology, {S}econd
  {E}dition}.
\newblock Chapman \& Hall/CRC Mathematical and Computational Biology. Taylor \&
  Francis.

\bibitem[\protect\citeauthoryear{Xie, Kurtek, Bharath, and Sun}{Xie
  et~al.}{2017}]{X_Weiyi_2017}
Xie, W., S.~Kurtek, K.~Bharath, and Y.~Sun (2017).
\newblock A geometric approach to visualization of variability in functional
  data.
\newblock {\em Journal of the American Statistical Association\/}~{\em
  112\/}(519), 979--993.

\bibitem[\protect\citeauthoryear{Xu and Tian}{Xu and Tian}{2015}]{D_Xu_2015}
Xu, D. and Y.~Tian (2015).
\newblock A comprehensive survey of clustering algorithms.
\newblock {\em Annals of Data Science\/}~{\em 2\/}(2), 165--193.

\bibitem[\protect\citeauthoryear{Zhu, Li, Ma, Wang, Liu, Huang, Zhang, and
  Hu}{Zhu et~al.}{2018}]{G_Zhu_2018}
Zhu, G., X.~Li, J.~Ma, Y.~Wang, S.~Liu, C.~Huang, K.~Zhang, and X.~Hu (2018).
\newblock A new moving strategy for the sequential monte carlo approach in
  optimizing the hydrological model parameters.
\newblock {\em Advances in Water Resources\/}~{\em 114}, 164--179.

\end{thebibliography}


\begin{thebibliography}{}

\bibitem[\protect\citeauthoryear{Fitzhugh}{Fitzhugh}{1961}]{R_Fitzhugh_1961}
Fitzhugh, R. (1961).
\newblock Impulses and physiological states in theoretical models of nerve
  membrane.
\newblock {\em Biophysical Journal\/}~{\em 1\/}(6), 445--466.

\bibitem[\protect\citeauthoryear{Hodgkin and Huxley}{Hodgkin and
  Huxley}{1952}]{A_Hodgkin_1952}
Hodgkin, A.~L. and A.~F. Huxley (1952).
\newblock A quantitative description of membrane current and its application to
  conduction and excitation in nerve.
\newblock {\em Journal of Physiology\/}~{\em 117\/}(4), 500--544.

\bibitem[\protect\citeauthoryear{Nagumo, Arimoto, and Yoshizawa}{Nagumo
  et~al.}{1962}]{J_Nagumo_1962}
Nagumo, J., S.~Arimoto, and S.~Yoshizawa (1962).
\newblock An active pulse transmission line simulating nerve axon.
\newblock {\em Proceedings of the Institute of Radio Engineers\/}~{\em
  50\/}(10), 2061--2070.

\bibitem[\protect\citeauthoryear{Ramsay, Hooker, Campbell, and Cao}{Ramsay
  et~al.}{2007}]{J_Ramsay_2007}
Ramsay, J.~O., G.~Hooker, D.~Campbell, and J.~Cao (2007).
\newblock Parameter estimation for differential equations: A generalized
  smoothing approach.
\newblock {\em Journal of the Royal Statistical Society: Series B\/}~{\em
  69\/}(5), 741--796.

\end{thebibliography}


\end{document}


\maketitle

\section{Simulation: Clustering of Vector-valued Functions}\label{supsec1}

\begin{table}[!t]
	\begin{center}
		\caption{Average adjusted Rand indices (with standard deviations in parentheses), computed across 50 replicates, for (a) elastic $k$-means, (b) KMA, (c) $k$-means on unaligned, discretized functions, and (d) $k$-means on aligned, discretized functions. Best performance is highlighted in bold.}
		\label{table:sim2RandIndex}
		\begin{tabular}{*{7}{|c}|}
			\hline
			$N$ & $\tau$ & $K$ & (a) & (b) & (c) & (d) \\
			\hline
			\multirow{4}{*}{120} & \multirow{2}{*}{0.05} & 2 & \textbf{1.0 (0)} & 0.064 (0.090) & 0.0020 (0.015) & 1.0 (0) \\
			& & 3 & \textbf{1.0 (0)} & 0.16 (0.098) & 0.022 (0.031) & 1.0 (0) \\
			\cline{2-7}
			& \multirow{2}{*}{0.1} & 2 & \textbf{1.0 (0)} & 0.097 (0.16) & 0.0029 (0.015) & 0.99 (0.019) \\
			& & 3 & \textbf{1.0 (0)} & 0.16 (0.11) & 0.028 (0.035) & 0.98 (0.024) \\
			\hline
			\multirow{4}{*}{240} & \multirow{2}{*}{0.05} & 2 & \textbf{1.0 (0)} & 0.12 (0.18) & 0.0034 (0.012) & 1.0 (0) \\
			& & 3 & \textbf{1.0 (0)} & 0.16 (0.089) & 0.010 (0.017) & 1.0 (0) \\
			\cline{2-7}
			& \multirow{2}{*}{0.1} & 2 & \textbf{1.0 (0)} & 0.11 (0.21) & 0.0013 (0.0068) & 0.99 (0.013) \\
			& & 3 & \textbf{1.0 (0)} & 0.19 (0.089) & 0.014 (0.020) & 0.99 (0.011) \\
			\hline			
		\end{tabular}
	\end{center}
\end{table}

\begin{figure}[!t]
	\centering
	\begin{tabular}{|c c|}
		\hline
		Dimension 1 & Dimension 2 \\
		\includegraphics{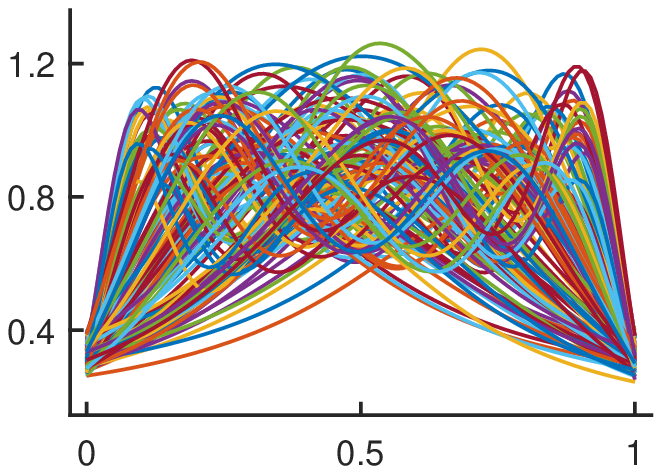} & \includegraphics{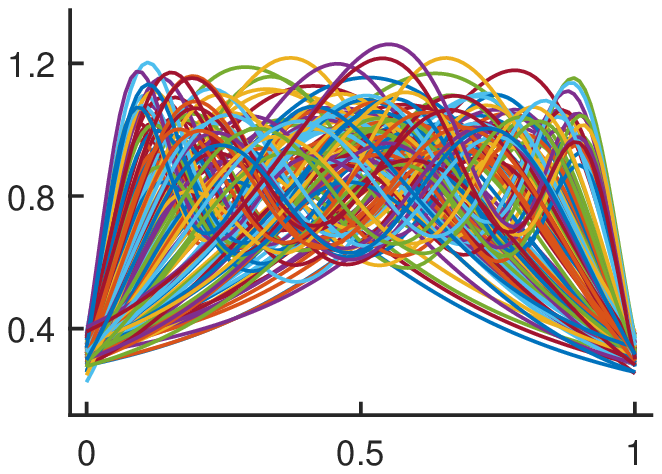} \\
		\hline
	\end{tabular}
	\caption{Spaghetti plot of one of the simulated datasets in Simulation 2, with sample size of 120, $\tau$ = 0.1, and number of true clusters equal to 3.}
	\label{fig:sim2spag}
\end{figure}

We present an additional simulation as an extension to the simulation that was carried out in the paper.
In this simulation, we consider samples of 2-dimensional vector-valued functions $f: [0,1] \rightarrow \mathbb{R}^2$, where the $i^{\text{th}}$ function before warping is $g_i(t) = (g_{i1}(t), g_{i2}(t))$. Let $b_{kl}$ denote the number of peaks in $g_{il}$ for the function in the $k^{\text{th}}$ cluster, and we set $b_{11}=2, b_{12} = 1, b_{21}=1, b_{22}=2, b_{31}=2, b_{32}=2$. Then,
\[
g_{il}(t) = \sum_{j = 1}^{b_{p_il}} z_{ijl}\phi\left(t;\frac{2b_{p_il}-1}{2b_{p_il}}, \frac{1}{3b_{p_il}}\right) , ~~t \in [0, 1],~~ l \in \{1,2\},
\]
where $\phi$ and $p_i$ are defined in the same way as in Simulation 1 in the paper, and $z_{ijl} \stackrel{iid}{\sim} N(1, \tau)$. The observed $f_i$ is obtained by applying the same random warping to $g_i$ as in Simulation 1. We vary the true number of clusters $K^*$ from 1 to 3, and again choose sample sizes $N$ to be either 120 or 240 and $\tau$ to be either 0.05 or 0.1. 

\begin{figure}[!t]
	\centering
	\begin{tabular}{|c c|c c|}
		\hline
		\multicolumn{2}{|c|}{(a)} & \multicolumn{2}{c|}{(b)} \\ 
		\includegraphics{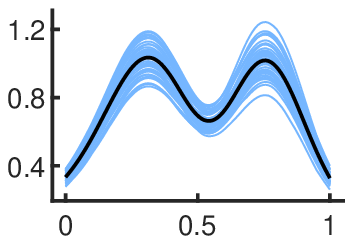} & \includegraphics{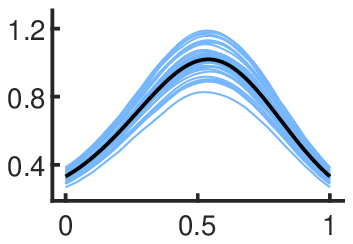} & \includegraphics{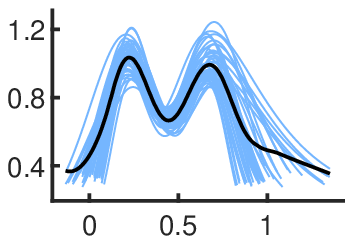} & \includegraphics{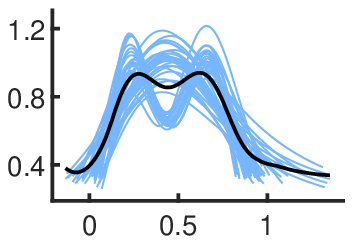} \\
		\includegraphics{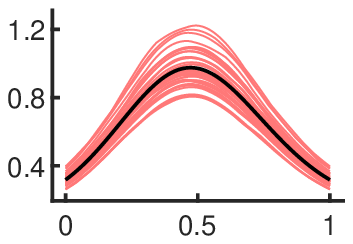} & \includegraphics{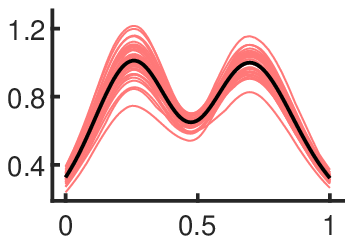} & \includegraphics{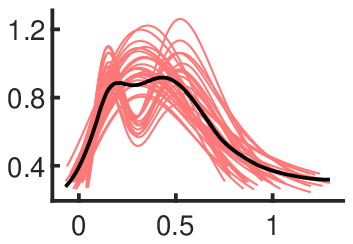} & \includegraphics{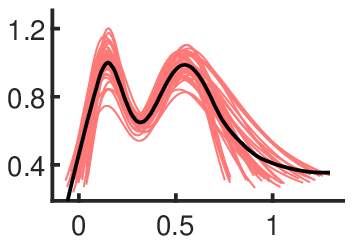} \\
		\includegraphics{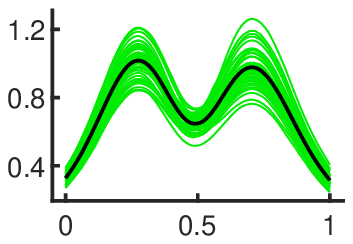} & \includegraphics{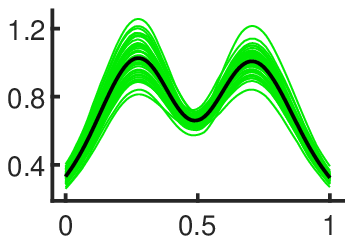} & \includegraphics{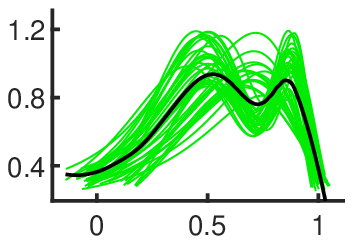} & \includegraphics{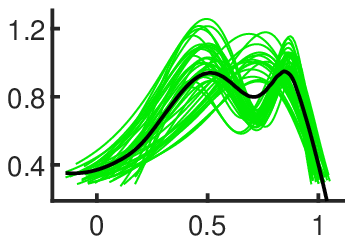} \\
		\hline
		\multicolumn{2}{|c|}{(c)} & \multicolumn{2}{c|}{(d)}  \\
		\includegraphics{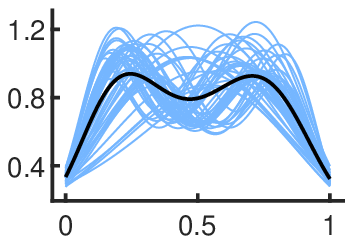} & \includegraphics{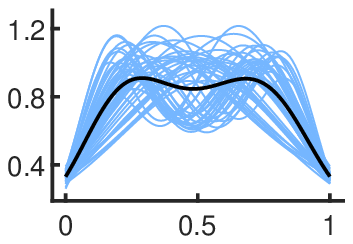} & \includegraphics{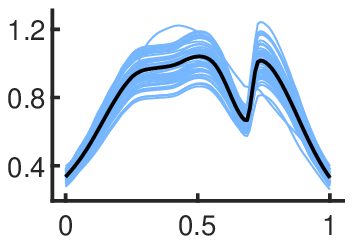} & \includegraphics{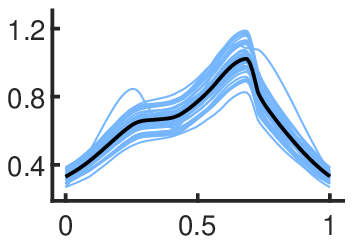} \\
		\includegraphics{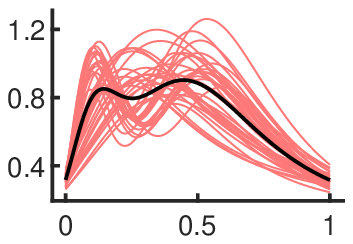} & \includegraphics{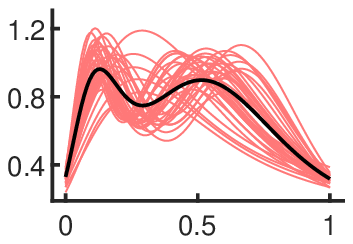} & \includegraphics{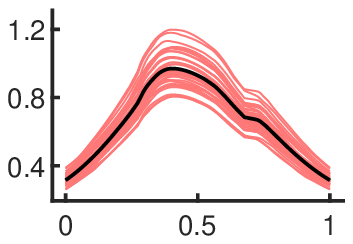} & \includegraphics{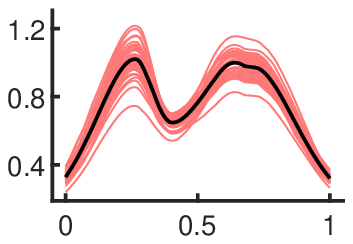} \\
		\includegraphics{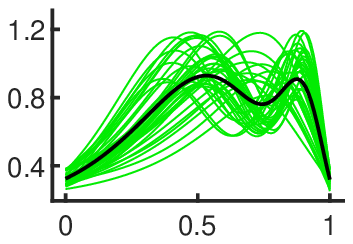} & \includegraphics{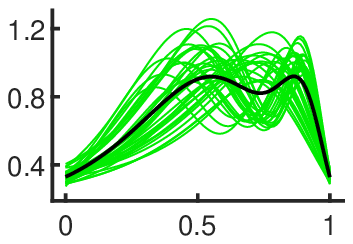} & \includegraphics{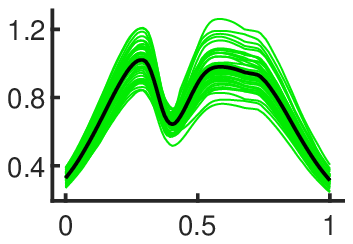} & \includegraphics{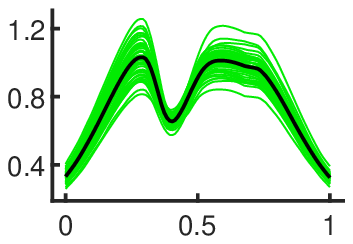} \\
		\hline
	\end{tabular}
	\caption{Clustering of functions in Figure \ref{fig:sim2spag} via different methods: (a) elastic $k$-means, (b) KMA, (c) $k$-means on unaligned, discretized functions, and (d) $k$-means on aligned, discretized functions. The left column of each block shows the first dimension and the right column shows the second dimension. The template function for each cluster is shown in black.}
	\label{fig:sim2results}
\end{figure}

\begin{figure}[!t]
	\centering
	\begin{tabular}{|c| c c c|}
		\hline
		\includegraphics{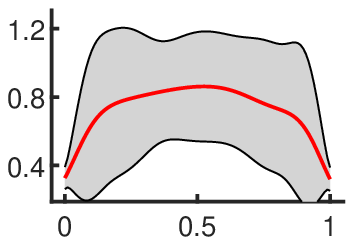} & \includegraphics{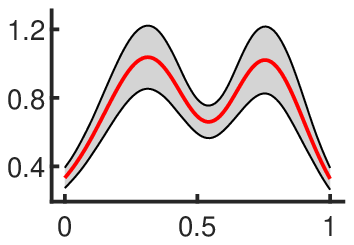} & \includegraphics{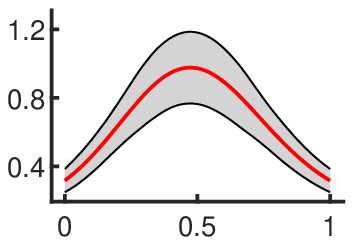} & \includegraphics{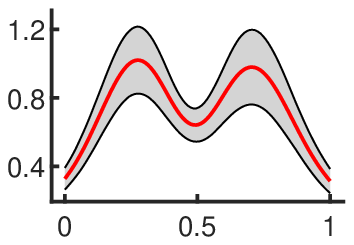} \\
		\includegraphics{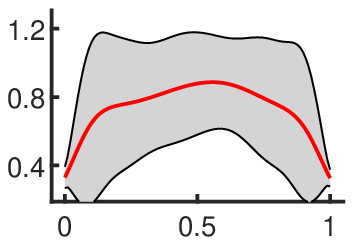} & \includegraphics{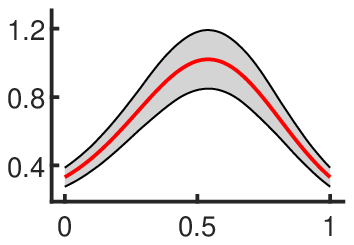} & \includegraphics{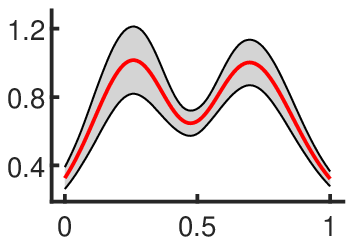} & \includegraphics{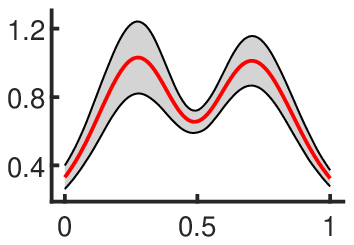} \\
		\hline
	\end{tabular}
	\caption{Pointwise summaries (mean $\pm$ 2 standard deviations) of vector valued functions in Figure \ref{fig:sim2spag} before (first column) and after clustering and alignment via elastic $k$-means (right three columns). The first and second rows correspond to the first and second dimension of the functions, respectively.}
	\label{fig:sim2bands}
\end{figure}

\begin{figure}[!t]
	\centering
	\begin{tabular}{|c|c|c|c|c|}
		\hline
		$N$ & $\tau$ & $K^*$=1 & $K^*$=2 & $K^*$=3\\
		\hline
		\multirow{2}{*}{120} & 0.05 & \includegraphics{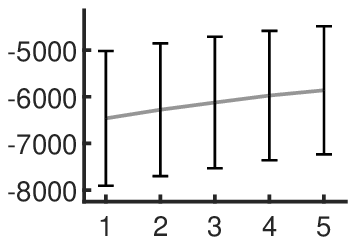} & \includegraphics{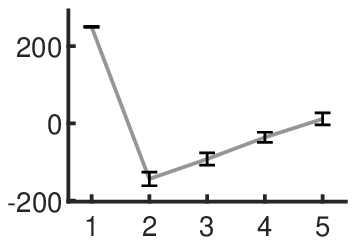} &
		\includegraphics{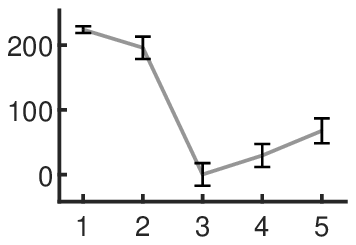} \\
		\cline{2-5}
		& 0.1 & 
		\includegraphics{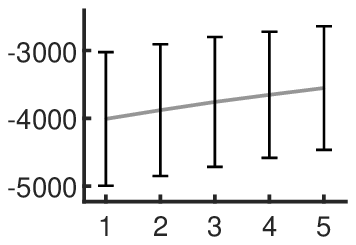} &
		\includegraphics{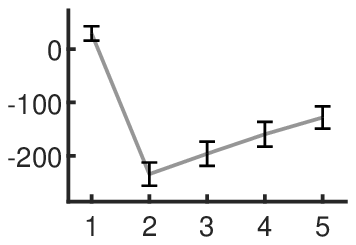} &
		\includegraphics{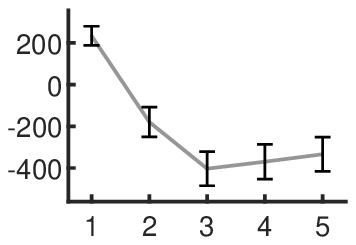}  \\
		\hline
		\multirow{2}{*}{240} & 0.05 & \includegraphics{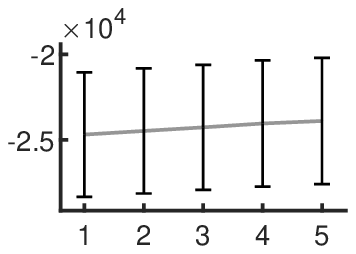} & \includegraphics{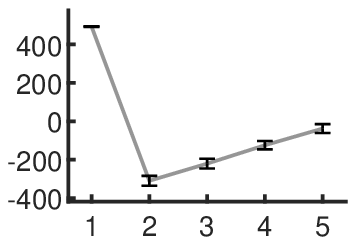} &
		\includegraphics{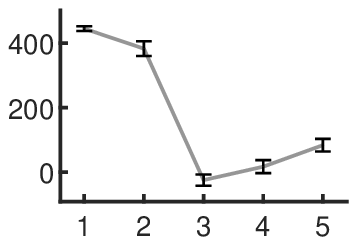}  \\
		\cline{2-5}
		& 0.1 & 
		\includegraphics{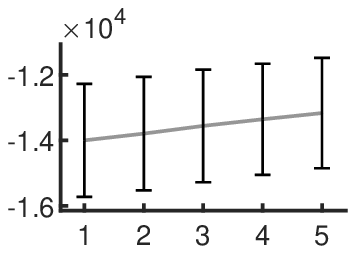} &
		\includegraphics{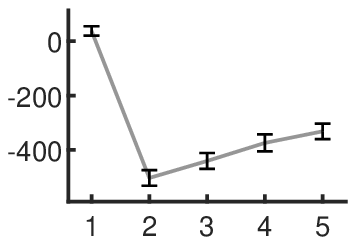} &
		\includegraphics{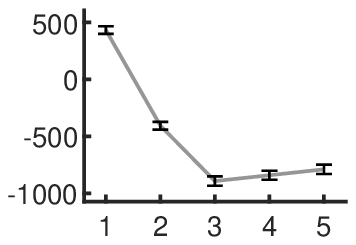}  \\
		\hline
	\end{tabular}
	\caption{Selection of $K$ using the proposed BIC criterion. Error bars show one standard deviation around the average BIC across 50 replicates.}
	\label{fig:sim2bic}
\end{figure}

Results summarized in Supplementary Table \ref{table:sim2RandIndex} and Supplementary Figures \ref{fig:sim2spag}-\ref{fig:sim2bic} are similar to those of Simulation 1 in the main paper with the following differences. While the clustering accuracy of elastic $k$-means and KMA is similar to what was observed for one-dimensional functions in Simulation 1, the performance of multivariate $k$-means applied to unaligned, discretized functions is closer to that of random clustering. We also notice that the accuracy of the $k$-means algorithm applied to aligned, discretized functions does not show as large a decrease in performance as in Simulation 1 in the main article when the within-cluster amplitude variability was increased. To provide some intuition, let us denote the functions $f$ by $f^{21}$, $f^{12}$, $f^{22}$, where the first number in the superscript identifies the number of peaks in the first dimension, and the second number identifies the number of peaks in the second dimension. When $K^* = 2$, the sample contains both $f^{21}$ and $f^{12}$. The second dimension of $f$ essentially corresponds to $K^* = 2$ of Simulation 1, but now we have one more dimension which doubles the available information for clustering as $f^{21}$ and $f^{12}$ differ in both dimensions. When $K^* = 3$, $f^{21}$ and $f^{12}$ distinguish themselves from $f^{22}$ in only one dimension. However, the ambiguity in peak matching is now reduced, because $f^{21}$ and $f^{12}$ both have at least one dimension with two peaks as in $f^{22}$, and there is only one way to align the two peaks in $f^{21}$ or $f^{12}$ to the two peaks of the corresponding dimension in $f^{22}$.

As shown in Supplementary Figure \ref{fig:sim2bands}, coordinate-wise summaries of given functions are much more meaningful after elastic $k$-means clustering. In particular, each clusters captures different amplitude structure in the given sample, which results in much more meaningful uncertainty bands. Supplementary Figure \ref{fig:sim2bic} again shows that the proposed model-based criterion to select the appropriate number of clusters performs well for vector valued functional data. As in Simulation 1 in the main article, the most difficult case arises when no clustering structure is present in the given sample.

\begin{figure}[!t]
	\begin{center}
		\begin{tabular}{|c|c|c|}
			\hline
		 (a)&(b)&(c)\\
		 \hline
		 \includegraphics[width=1.8in]{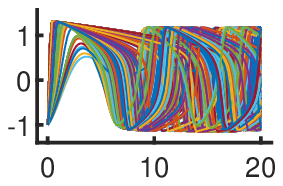} &
			\includegraphics[width=1.8in]{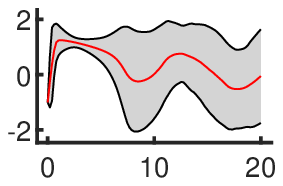} &
			\includegraphics[width=1.8in]{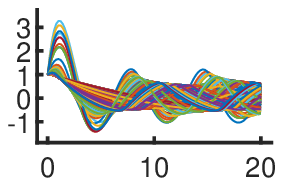} \\
			\hline
			(d)&(e)&(f)\\
			\hline
			\includegraphics[width=1.8in]{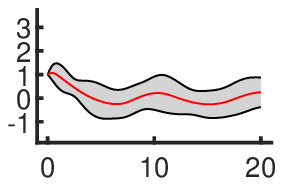} &
			\includegraphics[width=1.8in]{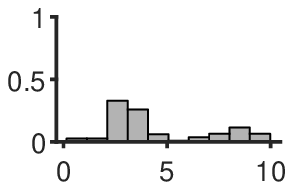} &
			\includegraphics[width=1.8in]{bic_nagumo_vr_\nagumosize_\thresh} \\
			\hline
		\end{tabular}
		\caption{(a)\&(c) Posterior samples of $v$ and $r$ functions, respectively. (b)\&(d) Pointwise summaries (mean $\pm$ 2 standard deviations) of the $v$ and $r$ posterior sample functions, respectively. (e) Histogram of posterior sample of $c$. (f) Optimal choice of the number of clusters via the proposed BIC.}
		\label{fig:nagumo}
	\end{center}
\end{figure}

\begin{figure}[!t]
	\begin{center}
		\begin{tabular}{|c c c c c c|}
			\hline
			\includegraphics[width=0.9in]{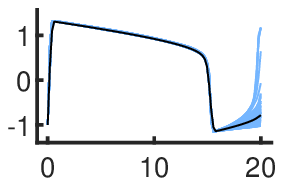} &
			\includegraphics[width=0.9in]{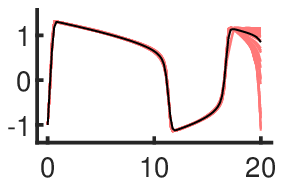} &
			\includegraphics[width=0.9in]{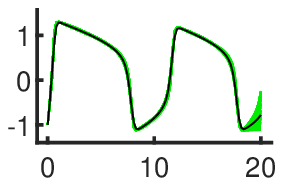} &
			\includegraphics[width=0.9in]{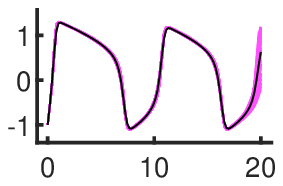} &
			\includegraphics[width=0.9in]{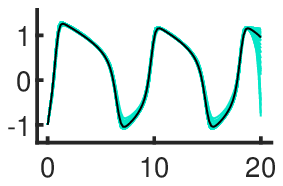} &
			\includegraphics[width=0.9in]{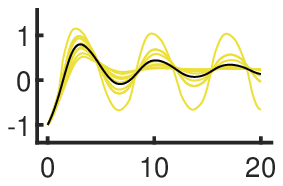} \\
			\includegraphics[width=0.9in]{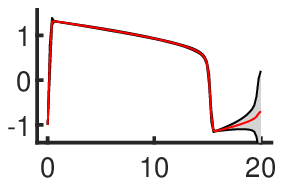} &
			\includegraphics[width=0.9in]{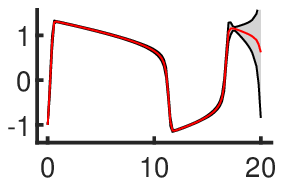} &
			\includegraphics[width=0.9in]{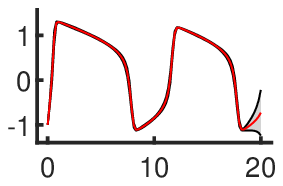} &
			\includegraphics[width=0.9in]{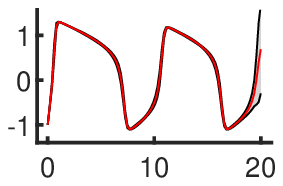} &
			\includegraphics[width=0.9in]{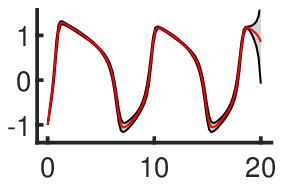} &
			\includegraphics[width=0.9in]{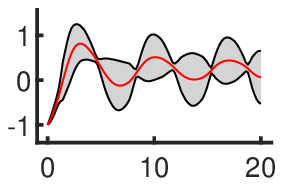} \\
			\hline
			\includegraphics[width=0.9in]{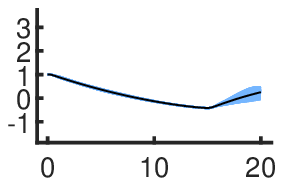} &
			\includegraphics[width=0.9in]{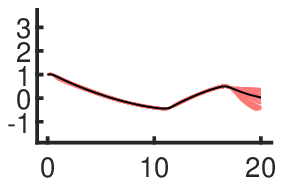} &
			\includegraphics[width=0.9in]{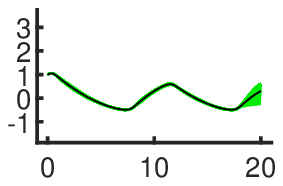} &
			\includegraphics[width=0.9in]{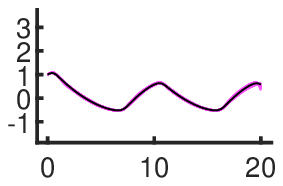} &
			\includegraphics[width=0.9in]{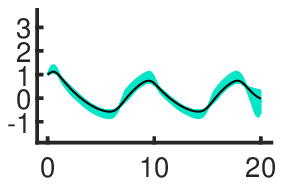} &
			\includegraphics[width=0.9in]{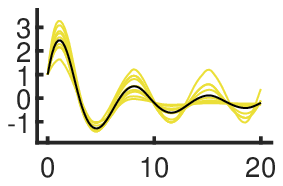} \\
			\includegraphics[width=0.9in]{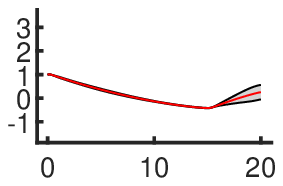} &
			\includegraphics[width=0.9in]{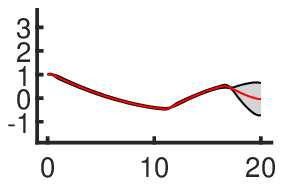} &
			\includegraphics[width=0.9in]{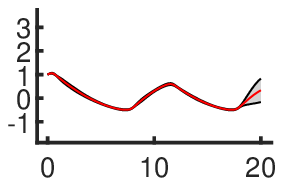} &
			\includegraphics[width=0.9in]{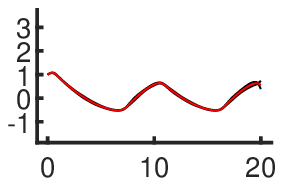} &
			\includegraphics[width=0.9in]{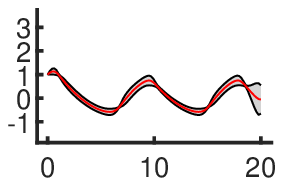} &
			\includegraphics[width=0.9in]{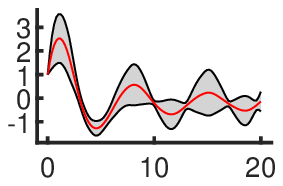} \\
			\hline
			\includegraphics[width=0.9in]{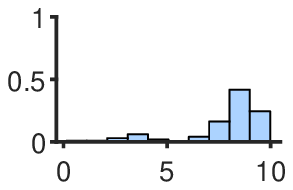} &
			\includegraphics[width=0.9in]{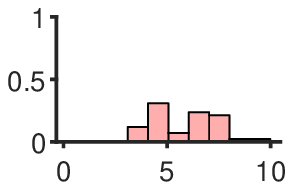} &
			\includegraphics[width=0.9in]{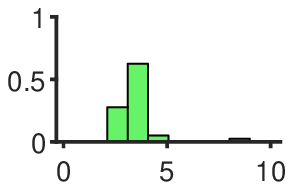} &
			\includegraphics[width=0.9in]{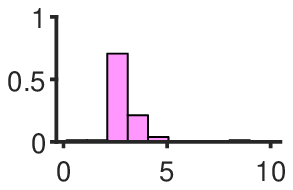} &
			\includegraphics[width=0.9in]{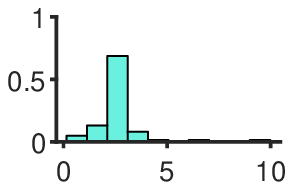} &
			\includegraphics[width=0.9in]{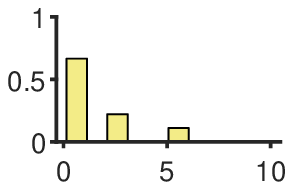} \\
			\hline
		\end{tabular}
		\caption{Result of elastic $k$-means applied to the FHN system posterior sample functions. Each column corresponds to a cluster. First row: Clustered and aligned $v$ functions. Second row: Pointwise summaries (mean $\pm$ 2 standard deviations) of the clustered and aligned $v$ functions shown in the first row. Third and fourth rows: Same as the first two rows, but for the $r$ functions. Last row: Histogram of $c$ corresponding to each cluster.}
		\label{fig:nagumo_frkma}
	\end{center}
\end{figure}

\section{Clustering Posterior Trajectories of the FitzHugh-Nagumo System}\label{supsec2}
We present an additional simulation that considers variation in neural data.
In particular, we consider the FitzHugh-Nagumo system (FHN, \citet{R_Fitzhugh_1961}, \citet{J_Nagumo_1962}).
The FHN system is a simplification of the Hodgekin-Huxley model \citep{A_Hodgkin_1952} for the excitation of neurons. The model is formulated via the system of ordinary differential equations:
\begin{equation*}
\begin{split}
\frac{dv}{dt} & = c(v - v^3 + r), \\
\frac{dr}{dt} & = -\frac{1}{c}(v - a + br),
\end{split}
\end{equation*}
where $v$, which stands for voltage, may be identified with membrane potential. Although $r$ may be interpreted as the recovery force that gradually diminishes $v$, it does not correspond to any specific physiological quantity that could be measured. The FHN model has been widely used to illustrate parameter estimation of dynamical systems from noisy observations (see, e.g.,  \citet{J_Ramsay_2007}). We consider a somewhat artificial setting where both $v$ and $r$ are observed with i.i.d. Gaussian noise. A dataset is simulated with $a = 0.2, b = 0.2, c = 3$ and initial values $v(0) = -1, r(0) = 1$ for $t \in [0, 20]$. The Gaussian noise has variance $\sigma^2 = 0.1$ for both $v$ and $r$. Assuming that $a$ and $b$ are known, we consider the problem of estimating $c$ from the the data using a Bayesian model with Inv-Exp$(1)$ prior on the error variance and Uniform$[0, 10]$ prior on $c$. Note that the marginal posterior over $c$ induces a posterior over the function $(v, r)$. We use MCMC to generate a sample of such posterior functions $(v, r)$ and apply our two-dimensional clustering approach as a means of posterior visualization. Panels (a) and (b) in Supplementary Figure \ref{fig:nagumo} show the marginal posterior samples of $v$ and their pointwise summaries. Panels (c) and (d) show the posterior samples of $r$ and their pointwise summaries. Panel (e) shows the histogram for the marginal posterior sample over $c$. Finally, panel (f) shows the proposed BIC, which indicates that the best number of clusters in this case is 6.

\begin{figure}[!t]
	\begin{center}
		\begin{tabular}{|c|c|c c|}
			\hline
			& (a) & (b) & (c) \\
			\hline
			(1) &
			\includegraphics{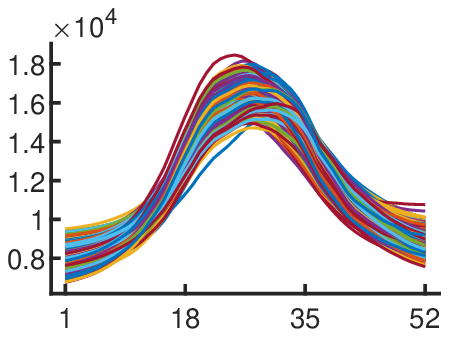} &
			\includegraphics{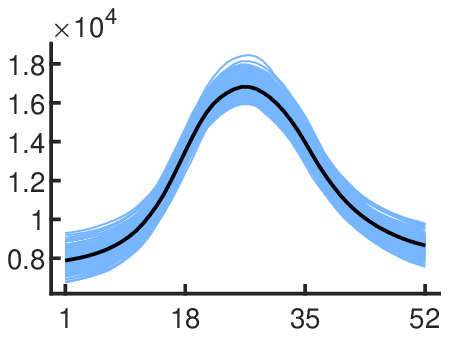} &
			\includegraphics{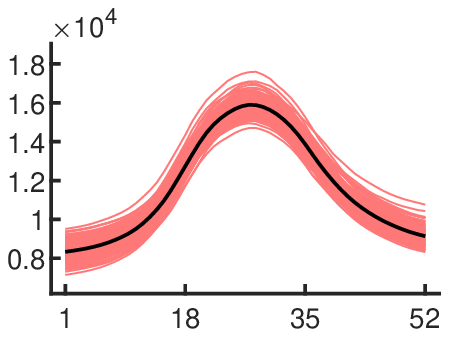} \\
			\hline
			(2) &
			\includegraphics{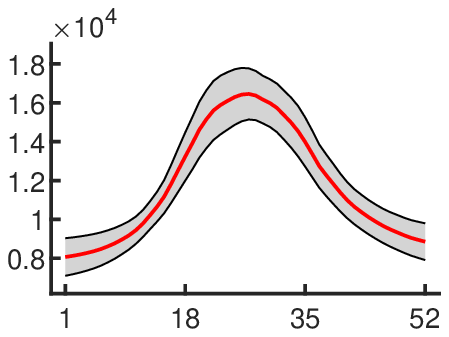} &
			\includegraphics{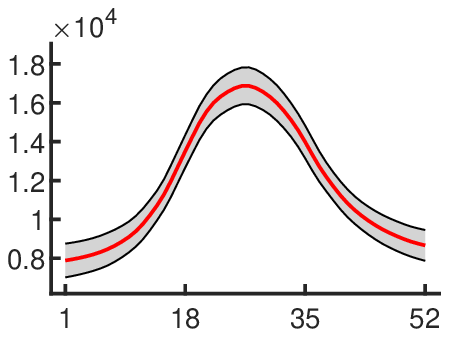} &
			\includegraphics{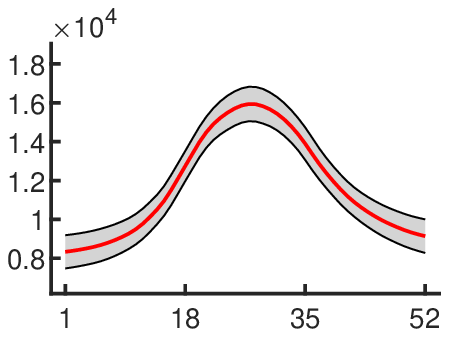} \\
			\hline
		\end{tabular}
		\caption{Application of elastic $k$-means clustering to 400 posterior draws of aggregated ARI trajectories for the year 2006-07. (1)(a) Original sample of functions. (1)(b)-(c) Two clusters (blue and red) of functions that were optimally aligned within each cluster (estimated cluster templates in black). The number of clusters was automatically determined via the proposed BIC. The blue cluster appears to have a higher mode than the red cluster. (2)(a)-(c) Pointwise summaries (mean $\pm$ two standard deviations) for the functions shown in (1)(a)-(c), respectively.}
		\label{fig:ari_2006}
	\end{center}
\end{figure}

\begin{figure}[!t]
	\begin{center}
		\begin{tabular}{|c|c|c|c|c|}
		    \hline
		    & (a) & (b) & (c) & (d) \\
			\hline
			(1) &
			\includegraphics[width=1.4in]{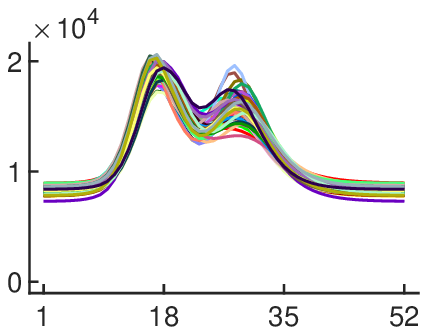} &
			\includegraphics[width=1.4in]{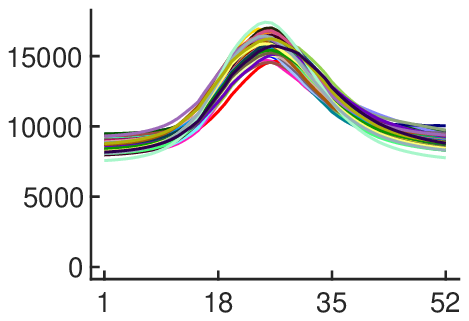} &
			\includegraphics[width=1.4in]{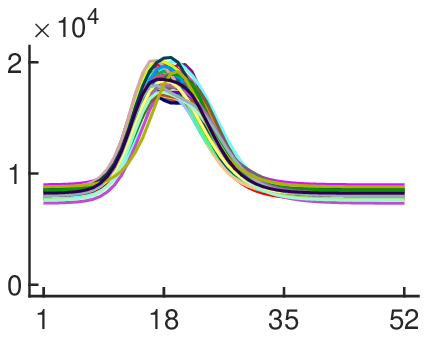} &
			\includegraphics[width=1.4in]{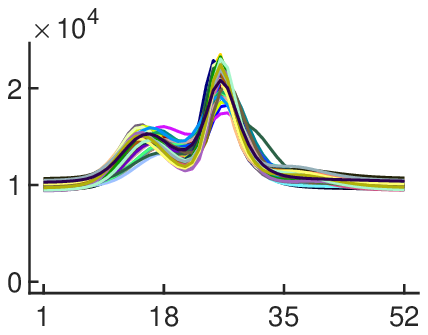}\\
			(2) &
			\includegraphics[width=1.4in]{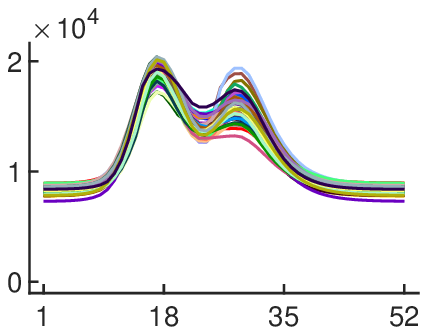} &
			\includegraphics[width=1.4in]{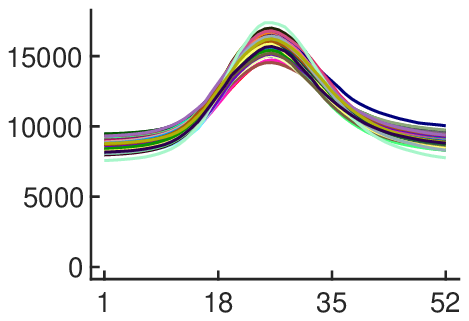} &
			\includegraphics[width=1.4in]{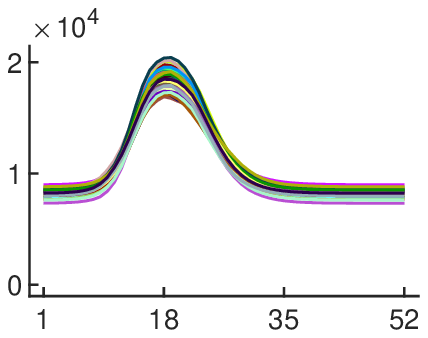} &
			\includegraphics[width=1.4in]{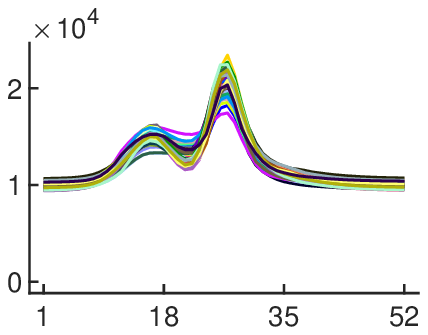}\\
			\hline
			(3) &
			\includegraphics[width=1.4in]{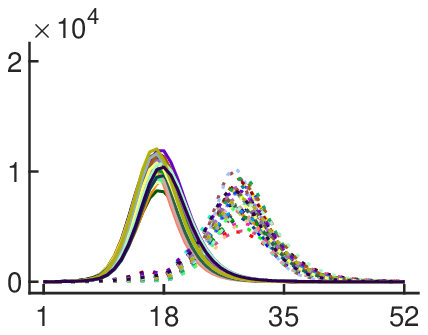} &
			\includegraphics[width=1.4in]{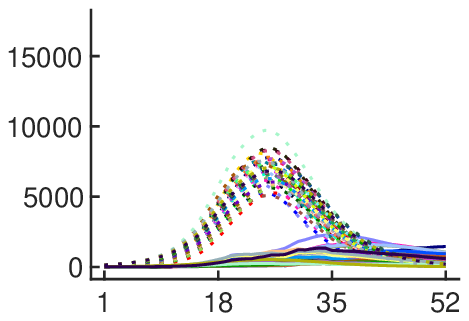} &
			\includegraphics[width=1.4in]{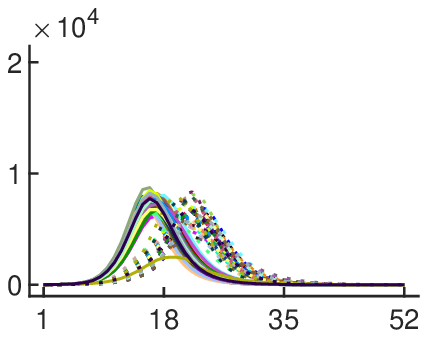} &
			\includegraphics[width=1.4in]{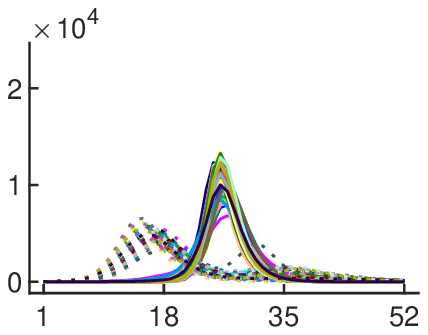}\\
			\hline
		\end{tabular}
		\caption{Multiple alignment results for posterior aggregated ARI trajectories for years without clustering structure (as suggested by the proposed BIC criterion). (a)-(d) Years 2003-04, 2004-05, 2005-06, 2007-08, respectively. Row (1): Functions before alignment. Row (2): Functions after multiple alignment. Row (3): Disaggregated influenza (solid line) and RSV (dotted line) trajectories . Each plot shows 50 functions sampled from the original 400 for clearer display.}
		\label{fig:years_wo_clusters}
	\end{center}
\end{figure}

\begin{figure}[!t]
	\begin{center}
		\begin{tabular}{|c|c|c c c|}
			\hline
			& (a) & (b) & (c) & (d) \\
			\hline
			(1) &
			\includegraphics[width=1.4in]{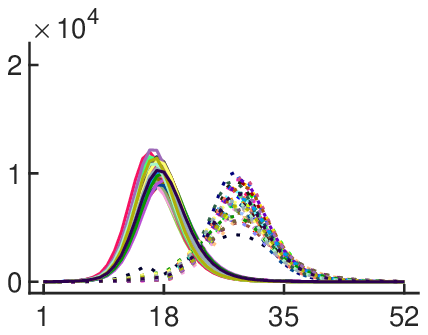} &
			\includegraphics[width=1.4in]{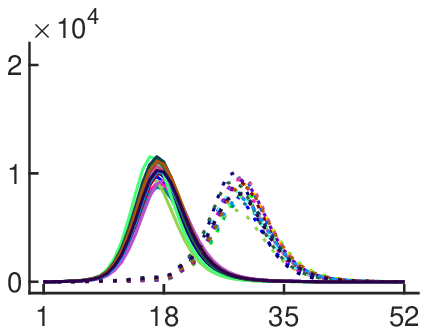} &
			\includegraphics[width=1.4in]{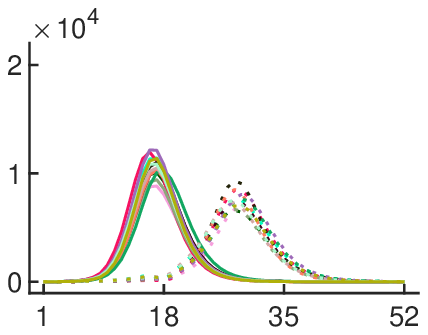} &
			\includegraphics[width=1.4in]{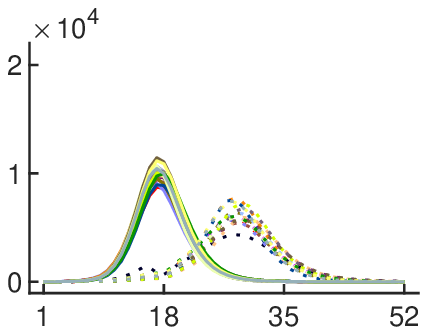} \\
			\hline
			(2) &
			\includegraphics[width=1.4in]{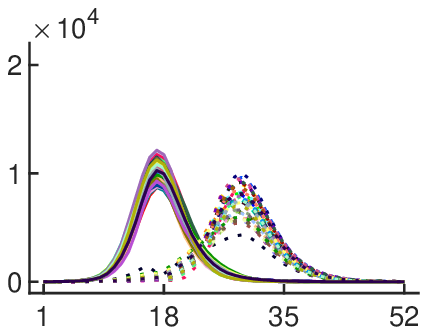} &
			\includegraphics[width=1.4in]{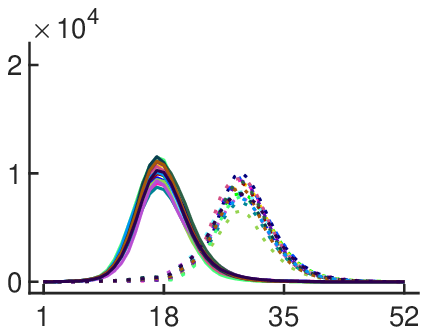} &
			\includegraphics[width=1.4in]{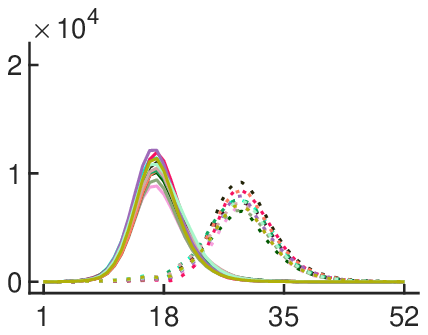} &
			\includegraphics[width=1.4in]{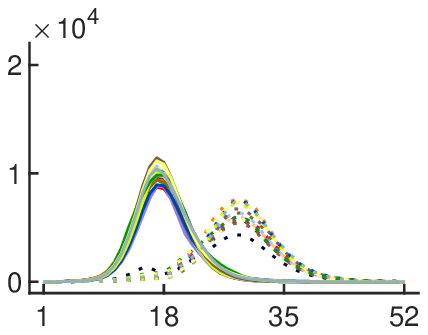} \\
			\hline
			(3) &
			\includegraphics[width=1.4in]{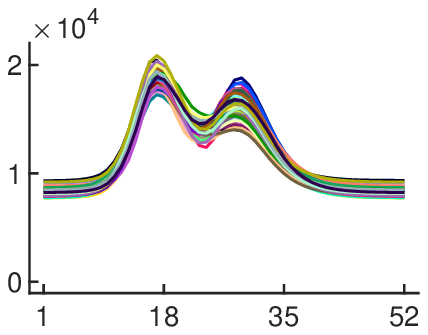} &
			\includegraphics[width=1.4in]{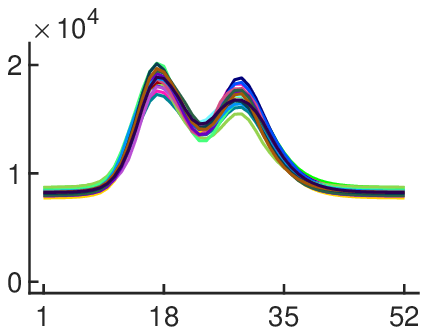} &
			\includegraphics[width=1.4in]{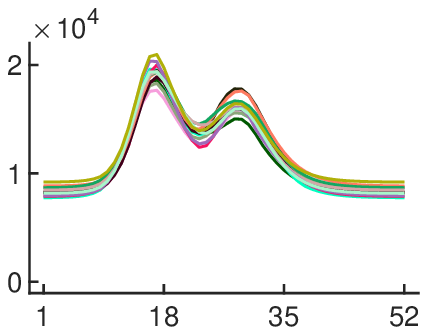} &
			\includegraphics[width=1.4in]{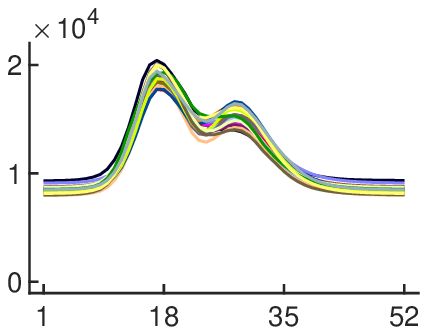} \\
			\hline
		\end{tabular}
		\caption{Elastic $k$-means clustering of disaggregated influenza (solid) and RSV (dotted) infection trajectories for the year 2003-04. (1)(a): Original sample of functions. (1)(b)-(d): Original sample of functions separated into three clusters corresponding to the three panels in Figure 11 in the main article. (2)(a): Original sample of functions after multiple alignment. (2)(b)-(d): Within cluster alignment via elastic $k$-means. (3): Influenza and RSV dimensions aggregated after alignment (with added background infection assumed to be constant over time). 50 functions sampled from the original 400 are plotted for clearer display.}
		\label{fig:supp_ari2d_2003_50}
	\end{center}
\end{figure}

\begin{figure}[!t]
	\begin{center}
		\begin{tabular}{|c|c|c c c|}
			\hline
			& (a) & (b) & (c) & (d) \\
			\hline
			(1) &
			\includegraphics[width=1.4in]{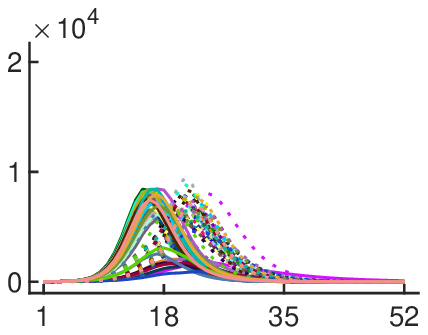} &
			\includegraphics[width=1.4in]{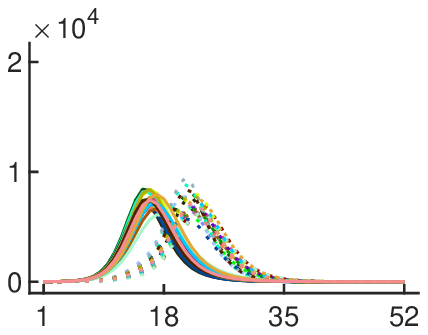} &
			\includegraphics[width=1.4in]{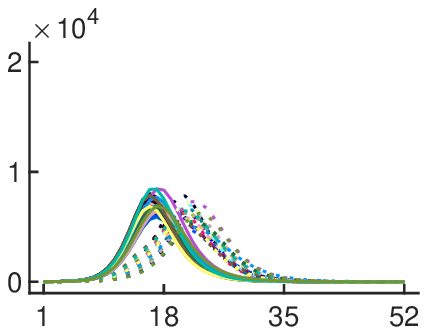} &
			\includegraphics[width=1.4in]{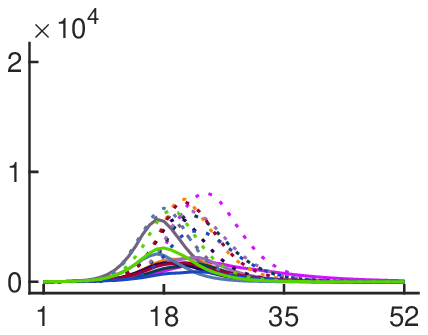} \\
			\hline
			(2) &
			\includegraphics[width=1.4in]{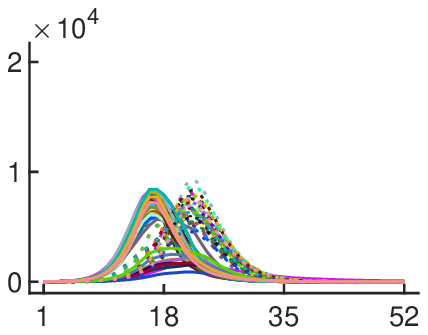} &
			\includegraphics[width=1.4in]{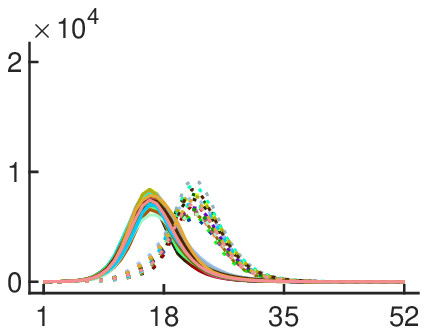} &
			\includegraphics[width=1.4in]{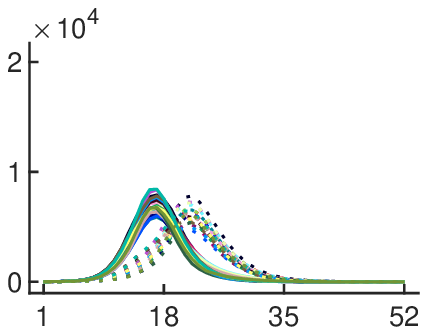} &
			\includegraphics[width=1.4in]{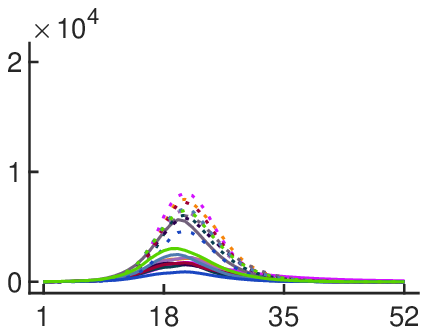} \\
			\hline
			(3) &
			\includegraphics[width=1.4in]{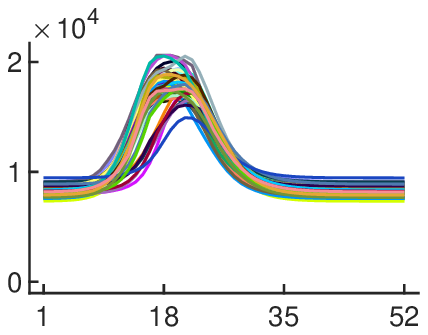} &
			\includegraphics[width=1.4in]{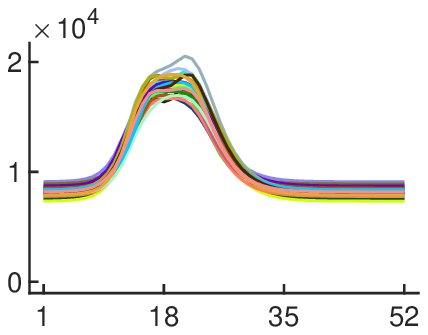} &
			\includegraphics[width=1.4in]{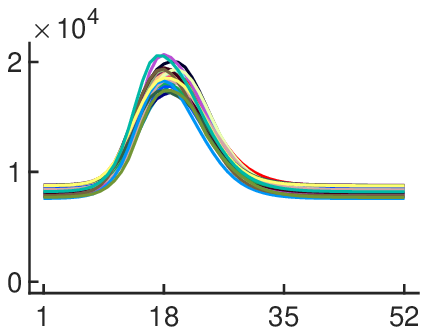} &
			\includegraphics[width=1.4in]{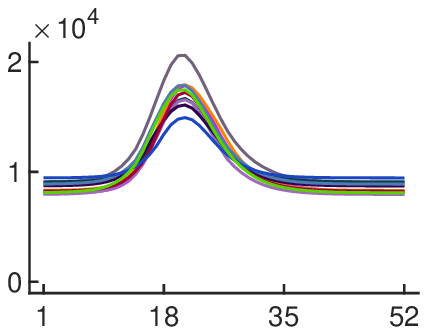} \\
			\hline
		\end{tabular}
		\caption{Elastic $k$-means clustering of disaggregated influenza (solid) and RSV (dotted) infection trajectories for the year 2005-06. (1)(a): Original sample of functions. (1)(b)-(d): Original sample of functions separated into three clusters corresponding to the three panels in Figure 12 in the main article. (2)(a): Original sample of functions after multiple alignment. (2)(b)-(d): Within cluster alignment via elastic $k$-means. (3): Influenza and RSV dimensions aggregated after alignment (with added background infection assumed to be constant over time). 50 functions sampled from the original 400 are plotted for clearer display (except for column(d) where all the functions in that cluster are plotted due to low prevalence.}
		\label{fig:supp_ari2d_2005_50}
	\end{center}
\end{figure}

\begin{figure}[!t]
	\begin{center}
		\begin{tabular}{|c|c|c c|}
			\hline
			& (a) & (b) & (c) \\
			\hline
			(1) &
			\includegraphics[width=1.4in]{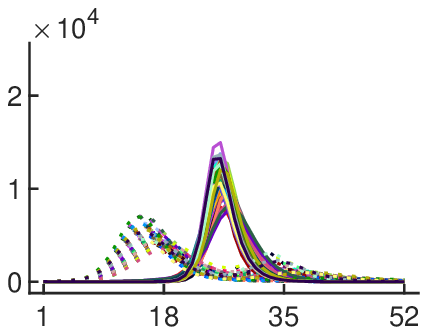} &
			\includegraphics[width=1.4in]{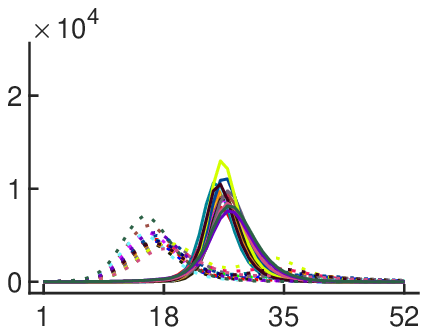} &
			\includegraphics[width=1.4in]{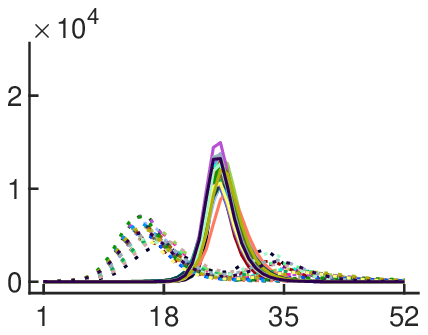} \\
			\hline
			(2) &
			\includegraphics[width=1.4in]{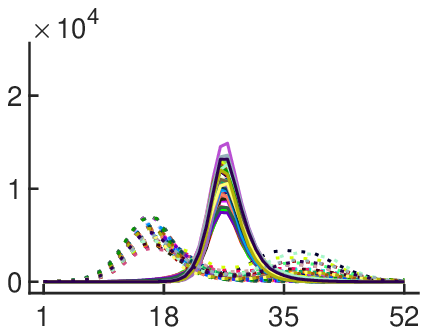} &
			\includegraphics[width=1.4in]{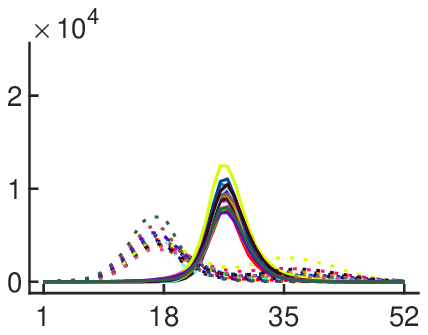} &
			\includegraphics[width=1.4in]{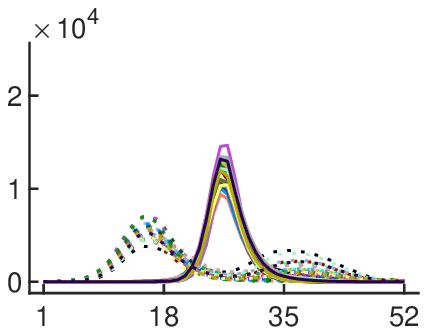} \\
			\hline
			(3) &
			\includegraphics[width=1.4in]{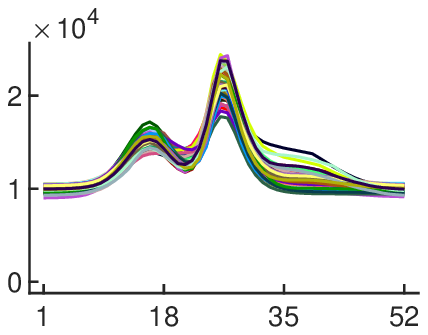} &
			\includegraphics[width=1.4in]{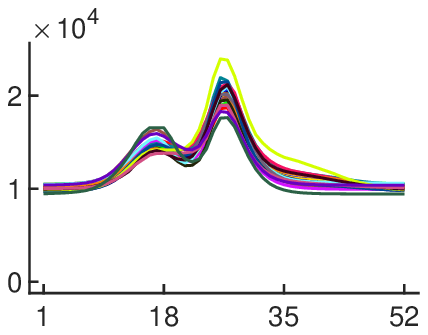} &
			\includegraphics[width=1.4in]{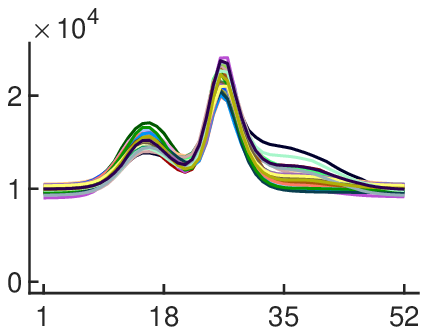} \\
			\hline
		\end{tabular}
		\caption{Elastic $k$-means clustering of disaggregated influenza (solid) and RSV (dotted) infection trajectories for the year 2007-08. (1)(a): Original sample of functions. (1)(b)-(d): Original sample of functions separated into two clusters corresponding to the two panels in Figure 13 in the main article. (2)(a): Original sample of functions after multiple alignment. (2)(b)-(d): Within cluster alignment via elastic $k$-means. (3): Influenza and RSV dimensions aggregated after alignment (with added background infection assumed to be constant over time). 50 functions sampled from the original 400 are plotted for clearer display.}
		\label{fig:supp_ari2d_2007_50}
	\end{center}
\end{figure}

\begin{figure}[!t]
	\begin{center}
		\begin{tabular}{|c|c|c|}
		    \hline
		    (a) & (b) & (c) \\
			\includegraphics{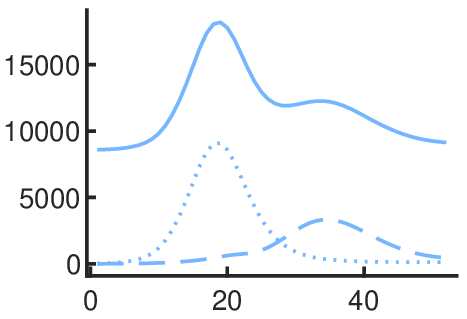} &
			\includegraphics{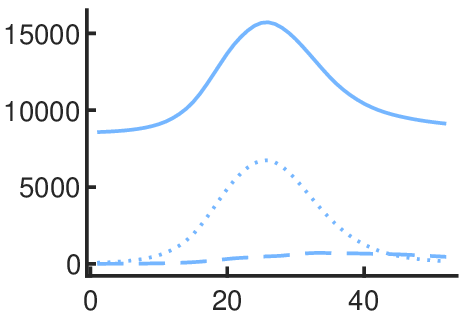} &
			\includegraphics{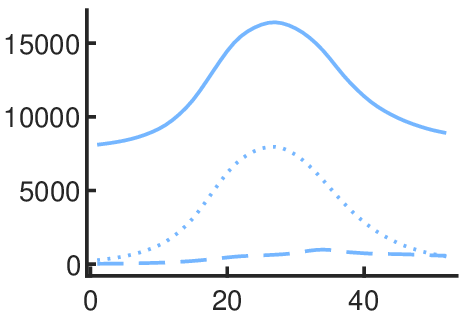} \\
			\hline
		\end{tabular}
		\caption{Templates after multiple alignment of 400 posterior draws of disaggregated influenza (dashed) and RSV (dotted) infection trajectories for years (a) 2002-03, (b) 2004-05 and (c) 2006-07 where no clustering structure was detected (as indicated by the proposed BIC). The solid function corresponds to an aggregated template based on the disaggregated influenza and RSV templates (with average background infections added).}
		\label{fig:ari_2d_1cl}
	\end{center}
\end{figure}

\begin{figure}[!t]
	\begin{center}
		\begin{tabular}{|c|c|c|c|}
		    \hline
		    & (a) & (b) & (c) \\
			\hline
			(1) &
			\includegraphics[width=1.4in]{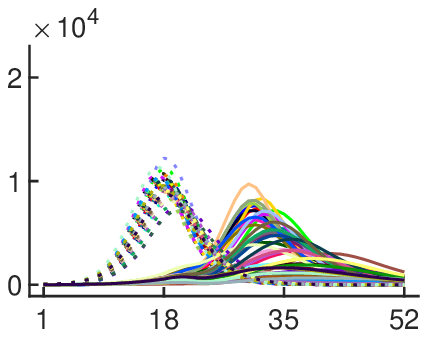} &
			\includegraphics[width=1.4in]{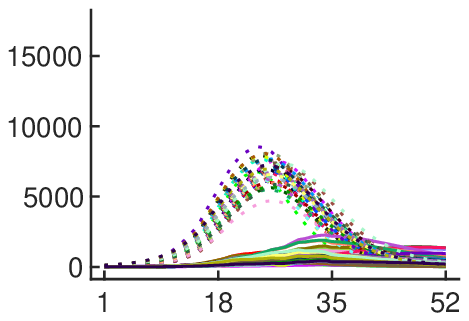} &
			\includegraphics[width=1.4in]{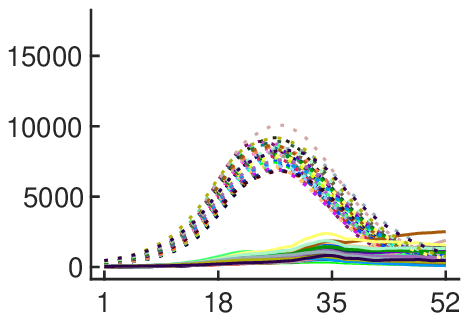} \\
			\hline
			(2) &
			\includegraphics[width=1.4in]{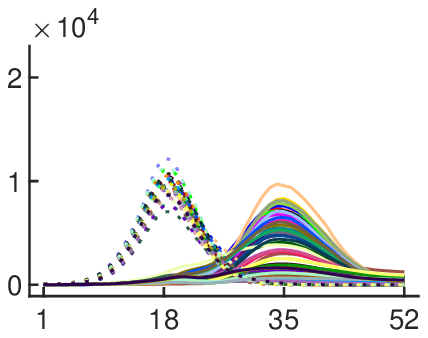} &
			\includegraphics[width=1.4in]{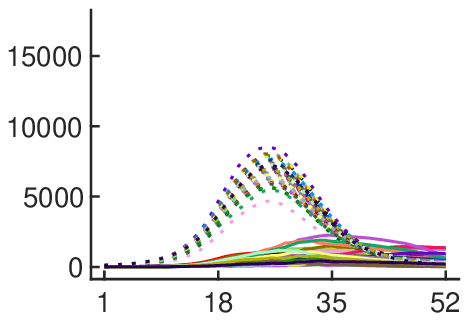} &
			\includegraphics[width=1.4in]{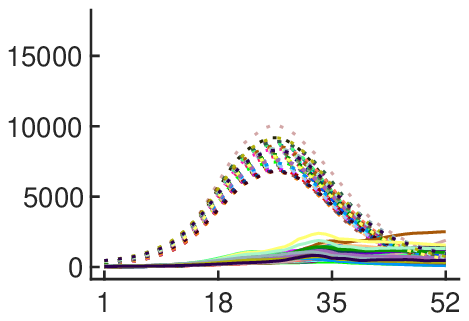} \\
			\hline
			(3) &
			\includegraphics[width=1.4in]{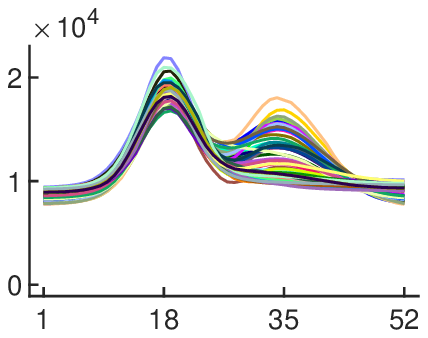} &
			\includegraphics[width=1.4in]{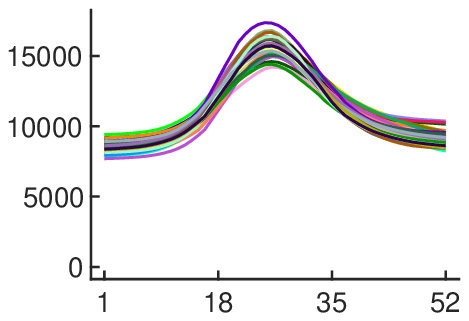} &
			\includegraphics[width=1.4in]{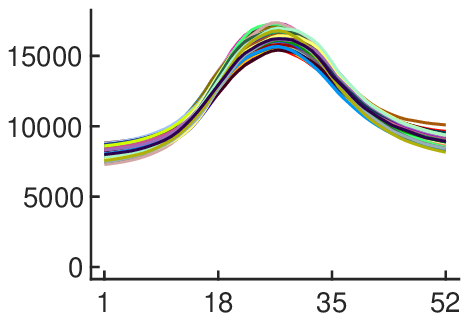}\\
			\hline
		\end{tabular}
		\caption{Multiple alignment results for posterior disaggregated ARI trajectories for years without clustering structure (as suggested by the proposed BIC criterion). (a)-(c) Years 2002-03, 2004-05, 2006-07, respectively. Row (1): Original disaggregated functions before alignment with influenza (solid) and RSV (dotted) shown together. Row (2): Disaggregated functions after multiple alignment. Row (3): Influenza and RSV dimensions aggregated after alignment (with added background infection assumed to be constant over time). Each plot shows 50 functions sampled from the original 400 for clearer display.}
		\label{fig:2d_years_wo_clusters}
	\end{center}
\end{figure}

We used the proposed BIC to estimate the number of clusters in the posterior sample, trying $K = 1, ..., 6$ with $\rho=0.95$. After a plateau at $K = 1,2,3$, the BIC decreases as $K$ increases. The decrease in BIC from $K=5$ to $K=6$ is small, but we choose $K=6$ as the optimal number of clusters in this case. The within-cluster alignments and pointwise summaries are shown in Supplementary Figure \ref{fig:nagumo_frkma} (each column corresponds to one of the six clusters). Elastic $k$-means separated the $(v,r)$ functions into clusters that are characterized by different numbers of peaks and valleys; this is especially evident in the first five clusters. We also identify a cluster, shown in the last column of the figure, that is characterized by function shapes that are very different from those in the first five clusters. Although the parameter $c$ is associated with the phase of the states $v$ and $r$, the association is not linear, as evidenced by the assignment of functions corresponding to similar $c$ values to different clusters. This suggests that functional clustering could not be replaced with simply clustering of this one-dimensional parameter. In particular, we discover much more structure in the posterior sample, by clustering the $(v,r)$ functions, than by simply clustering the parameter $c$.

\section{Additional Figures: Clustering and Visualization of Posterior ARI Trajectories}\label{supsec3}

In this section, we present additional figures that are referenced in the main paper. All of these figures correspond to our analysis of aggregated and disaggregated ARI trajectory posterior samples. Supplementary Figure 7 presents elastic $k$-means clustering results of aggregated ARI trajectories for infection year 2006-07. Supplementary Figure 8 presents multiple alignment results for aggregated ARI trajectories for infection years without clustering structure. Supplementary Figures 9-11 describe elastic $k$-means clustering results for disaggregated influenza and RSV posterior trajectories for infection years 2003-04, 2005-06 and 2007-08, respectively. Supplementary Figure 12 shows cluster templates for disaggregated influenza and RSC posterior trajectories for infection year 2002-03. Finally, Supplementary Figure 13 reports multiple multiple alignment results for disaggregated influenza and RSC posterior trajectories for infection years without clustering structure.

\clearpage
\bibliographystyle{chicago}
\bibliography{FRkmeansJournal}
